\documentclass[aps,twocolumn,nofootinbib,showpacs,final,balancelastpage,superscriptaddress]{revtex4}
\usepackage{graphicx}
\usepackage{tabularx}
\usepackage{amssymb}
\usepackage{amsmath}
\usepackage{amsbsy}
\usepackage{comment}
\usepackage{mathrsfs}
\usepackage{dsfont}
\usepackage{nicefrac}
\usepackage{wrapfig}
\usepackage{amsfonts}
\usepackage{url} 
\usepackage[nolist]{acronym}
\usepackage{nicefrac}
\usepackage{subfigure}
\usepackage{epsf,color,colordvi,pifont}
\usepackage{showkeys}

\usepackage{slashed}

\DeclareFontFamily{OT1}{pzc}{}
\DeclareFontShape{OT1}{pzc}{m}{it}{<-> s * [1.10] pzcmi7t}{}
\DeclareMathAlphabet{\mathpzc}{OT1}{pzc}{m}{it}

\begin{document}
\title{Non-linear Breit-Wheeler pair production  in  collisions of bremsstrahlung\\  $\gamma-$quanta and a tightly focussed  laser pulse}
\author{A. \surname{Golub}}
\email{Alina.Golub@uni-duesseldorf.de }
\affiliation{Institut f\"{u}r Theoretische Physik I, Heinrich-Heine-Universit\"{a}t D\"{u}sseldorf, Universit\"{a}tsstra\ss{e} 1, 40225 D\"{u}sseldorf, Germany.}
\author{S. \surname{Villalba-Ch\'avez}}
\email{selym@tp1.uni-duesseldorf.de}
\affiliation{Institut f\"{u}r Theoretische Physik I, Heinrich-Heine-Universit\"{a}t D\"{u}sseldorf, Universit\"{a}tsstra\ss{e} 1, 40225 D\"{u}sseldorf, Germany.}
\author{C. \surname{M\"{u}ller}}
\email{c.mueller@tp1.uni-duesseldorf.de}
\affiliation{Institut f\"{u}r Theoretische Physik I, Heinrich-Heine-Universit\"{a}t D\"{u}sseldorf, Universit\"{a}tsstra\ss{e} 1, 40225 D\"{u}sseldorf, Germany.}

\begin{abstract}
Experimental efforts toward the detection of the nonperturbative  strong-field regime of the Breit-Wheeler pair creation process plan to 
combine incoherent sources of GeV $\gamma$ quanta and the coherent fields  of tightly focussed optical laser pulses. This endeavour calls for a 
theoretical understanding of how the pair yields depend on the applied  laser field profile. We provide estimates for the number of produced 
pairs in a setup where the high-energy radiation is generated via  bremsstrahlung. Attention is paid to the role of the transversal and 
longitudinal focussing of the laser field, along with the incorporation  of a Gaussian pulse envelope. We compare our corresponding results with 
predictions from plane-wave models and determine the parameters of  focused laser pulses which maximize the pair yield at fixed pulse 
energy. Besides, the impact of various super-Gaussian profiles for the  laser pulse envelope and its transverse shape is discussed.
\end{abstract}

\keywords{Breit-Wheeler pair creation.}

\date{\today}

\maketitle

\section{Introduction}

Materializing  quantum vacuum fluctuations   into real electron-positron pairs  from  collisions of photons is among the  iconic predictions of quantum electrodynamics that  support the  contemporary perception  of the quantum vacuum  as  a  source of  nonlinear  electromagnetic interactions.  In fact,  theoretical studies that followed Breit and Wheeler's pioneering  work on the linear pair production by two photons \cite {BreitWheeler}  revealed  creation channels $\gamma^\prime+n\gamma\to e^-+e^+$  in which $n$ background photons  could be  absorbed simultaneously in the course of a single  pair creation event \cite{Reiss1,RitusReview,NikishovRitus,NikishovRitus2,baier,Reiss}.  This nonlinear landscape  is expected  to  occur  in  both  the perturbative weak-field  ($\xi \ll 1$) and   nonperturbative  strong-field  ($\xi\gg1$) regimes  ruled by  the  laser  intensity   parameter $\xi=\vert e\mathcal{E}_0\vert/(m\omega)$,  characterized by  the laser  frequency  $\omega$ and   peak  field strength   $\mathcal{E}_0$.\footnote{Throughout this paper we use Lorentz-Heaviside units where $c=\hbar=\epsilon_0=1$. Besides, the  signature of the metric tensor is chosen with $\mathrm{diag}(g^{\mu \nu})=(1,-1,-1,-1)$.  The symbols $e$ and $m$ stand for the electron charge and mass, respectively.}  In the former scenario the partial  rates linked to  nonlinear events  $R_n\sim \xi^{2n}$  are suppressed  as the number of absorbed photons ($n>1$) grows. This  means that ---  in practice ---    nonlinear  Breit-Wheeler  reactions   in the field of a laser with $\xi<1$  are likely to take place with the absorption of few photons only.  The described scenario was  confirmed experimentally  by the  SLAC E-144 collaboration  \cite{Burke}.  Besides,  by accelerating   gold ions  to ultra-relativistic energies,   an  experimental validation   of  the linear channel ($n=1$) in collisions of quasi-real photons has been reported recently  \cite{STAR}.

In contrast to the perturbative  scenario, a large amount  of  laser  photons is predicted to be absorbed  when  a single pair is  produced  under the condition $\xi\gg1$.  So far this  highly nonlinear  regime  lacks  an experimental   observation, mainly because  the  total production rate  $R\sim \exp[-8/(3\kappa)]$, with $\kappa=2\omega^\prime \mathcal{E}_0/(m E_c)$ for counterpropagating beam geometry, is exponentially  suppressed   unless  the effective peak strength  $(\omega'/m)\mathcal{E}_0$ comes close to  the characteristic Schwinger  scale  $E_c=m^2/\vert e\vert\approx 1.3 \times 10^{16}\; \rm Vcm^{-1}$.  While a field $E_0$ as large as $E_c$ is yet inaccessible in the laboratory frame,  the  possibility of  producing   highly energetic  $\gamma$ radiation   ($\omega'\gtrsim\mathcal{O}(1)\; \rm GeV$)  via bremsstrahlung   combined   with both the current availability  of  multi-petawatt laser facilities  \cite{diPiazzaReview}   and   unprecedented detection techniques,  makes  the first  confirmation of  the nonperturbative  strong-field  regime of  the  Breit-Wheeler  pair creation process  come into reach.  As a consequence,  various experimental endeavors are being  planned  worldwide, including the projects E-320 at SLAC  \cite{Meuren2019} and LUXE at DESY \cite{LUXE,LUXE2} as well as the upcoming experiments at the Rutherford Appleton Laboratory \cite{Appleton} and   the  one  put forward by  the Center of Advanced Laser Applications (CALA) \cite{Salgado}. 

Clearly,  as the high-intensity lasers  involved  in the  listed collaborations  ($I\lesssim\mathcal{O}(10^{22})\; \rm Wcm^{-2}$) are tightly focussed  and  the production of   $\gamma$ quanta  through  bremsstrahlung generates a broad spectrum,   attempts  in describing theoretically  the forthcoming measurements  require  to go beyond the  traditional treatments of pair production by a monoenergetic $\gamma$ beam and a plane-wave laser field, on which most of the  investigations carried out so far rely \cite{RitusReview,NikishovRitus,NikishovRitus2,baier,Reiss,Meuren:2014uia,Krajewska2012,Krajewska2014,Kaempfer2012,Jansen2013,VillalbaChavez:2012bb, Heinzl,Jansen2017, Grobe2018,Kaempfer2018,Kaempfer2020,Tang,X}. This situation renders theoretical studies  of the strong field pair  production process  in  realistic setups  a  subject of raising interest.  Although a  comprehensive theoretical framework including the aforementioned properties is far from being accomplished, progresses toward  understanding the role of the transversal focusing of the high-intensity  laser pulse  have been achieved at the fundamental level,  where a single $\gamma$ quantum of fixed energy intervenes  \cite{diPiazza2016, diPiazza2021,Riconda}.  Parallelly,  there have been efforts  for assessing  the impact of the spectral distribution of bremsstrahlung $\gamma$ quanta on the production of pairs  by adopting various models  for the  strong background laser field \cite{Hartin, Blackburn,Eckey}. 

\begin{figure}[ht]
\includegraphics[width=0.5\textwidth]{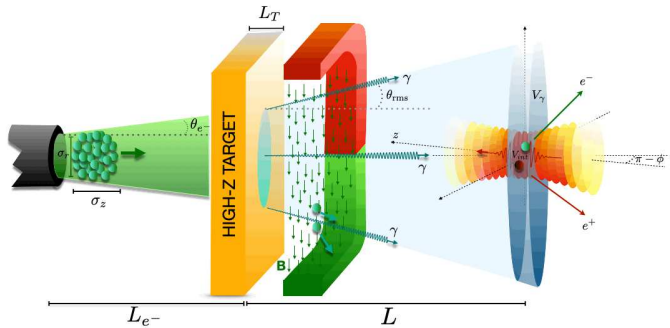}
\caption{\label{fig1} Sketch of  an  experimental  setup  put forward to create $e^-e^+$-pairs  from the collision of bremsstrahlung $\gamma$ quanta and a tightly focussed laser pulse  in the   nonperturbative strong-field regime  ($\xi\gg1,\;\kappa\sim 1$)  of the  Breit-Wheeler process. The high-$Z$ target responsible for the generation of the $\gamma$ radiation is  supposed to be thin. A magnet is located right after the target to deflect  primarily the electron flux that traverses  it.   The spreading  beam  of bremsstrahlung photons forms a cone coloured in blue.   For more details on the planed experiment we refer the reader to  Ref.~\cite{Salgado}.}
\end{figure}

The present  manuscript  aims to  provide first estimates for the expected pair yields by including simultaneously both the focusing of the high-intensity laser pulse as well as the frequency spectrum and spatial spreading of the bremsstrahlung $\gamma$ beam.   We consider a modern version  of  the setup proposed  originally by Reiss \cite{Reiss}, in which a high-intensity optical  laser  field   and  high energy $\gamma$ quanta  --- produced in the course of the interaction between  highly energetic incident electrons and a thin high-Z target --- collide, giving rise to  electron-positron pairs (see Fig.~\ref{fig1}).\footnote{A  similar  setup has been put forward   as an  alternative route for verifying the linear Breit-Wheeler reaction. The interested readers are   referred to Ref.~\cite{Golub}, where  the associated phenomenology is  discussed in details.}  The benchmark parameters  used  in our investigation are in correspondence with those envisaged  in Ref.~\cite{Salgado}, which guarantee the realization  of  the   nonperturbative strong-field regime  ($\xi \gg 1$,\; $\kappa \sim 1$) of the  Breit-Wheeler pair production process.  In this parameter regime,  a  Locally Constant Field Approximation (LCFA)  can be applied \cite{RitusReview,Meuren,King2015,King2019,King2020,King2021}.  Its use allows us to assess the role of the transversal and longitudinal focussing of the laser field along  with the incorporation of the Gaussian pulse envelope. We  show that the inclusion  of  the  spatial laser focusing   reduces the  pair production yield as compared to  scenarios relying on plane-wave pulses comprising the same pulse energy and establish   intensity-focusing parameters  for which  the number of production events is  optimized. Besides, the impact of choosing  super-Gaussian profiles for the pulse envelope and the transversal shape of the laser field is analysed separately.

The manuscript is organized as follows. In Sec.~\ref{GAs} we provide details of  the  theoretical framework   used in our analysis.  Particularly,  in  Sec.~\ref{bspe} various  aspects  of  the  bremsstrahlung $\gamma$ radiation are introduced  and some assumptions are adopted.  We then proceed with Sec.~\ref{laser}, where   some laser field profiles to be investigated are presented. Next,  in  Sec.~\ref{NCPs}    expressions for the production probability and the  number of  produced pairs per radiating electron are elucidated via  LCFA.  In parallel, an expression for the pair production rate  in a constant crossed field is derived for  the regime  where  the conditions  $\xi \gg1$  and  $\kappa \approx 1$ are fulfilled simultaneously.   The  formulae obtained in  Sec.~\ref{NCPs}  are afterwards exploited  to assess numerically the impact of the different laser field models.  The results of these evaluations are discussed in  Sec.~\ref{Results}.  While in Sec.~\ref{CDFM} comparisons between the constant crossed field,  plane-wave and paraxial field  models are carried out, Sec.~\ref{SGPE} is devoted to evaluate  effects linked to  super-Gaussian profiles. The role of different focal regions within the photo-production of  pairs is  studied in Sec. \ref{CFDFR}, whereas  the impact of the relation between tighter focus and higher intensity is investigated in Sec.~\ref{FEEFFF}.    Finally,  in Sec.~\ref{Conclusions},  we present our conclusion, whereas in the appendices details on  the space-dependent  quantum non-linearity parameter are given  and expressions for the electric field beyond the paraxial approximation are listed.   

\section{Theoretical approach\label{GAs}}

 The envisaged setup is split into two stages in which the generation of high-energy $\gamma$ radiation  and  the pair production processes  occur separately. While in the first stage, the $\gamma$ quanta are produced through  bremsstrahlung of an incident  highly collimated  beam of  ultrarelativistic  electrons, in the second stage a fraction of  them collides  with  a high-intensity laser pulse. This section is devoted to introduce the analytical tools that are used in the description of the strong-field  nonlinear Breit-Wheeler pair production as it may occur in the depicted  configuration (see  Fig.~\ref{fig1}). 

\subsection{Bremsstrahlung spectrum \label{bspe}}

In our numerical computations, we shall consider incident electron beams comprising several pC of total charge which have reached an energy $E_0$ of few GeV via  Laser Wake-Field Acceleration (LWFA). The spatial extent of such electron beams depends on the precise regime of acceleration. To be specific, we shall assume that the accelerating field  needed for exciting  the plasma wake is taken from the same   laser  source  which provides the tightly focussed laser pulse that drives the pair production process in the second stage, as it is experimentally planned \cite{Salgado}.  However,  in contrast to the latter pulse, the former  will be weakly focussed. The axial extension  of the electron beam $\sigma_z$  varies depending upon the plasma density $\mathpzc{n}_e<10^{18}\;\rm cm^{-3}$ and the parameters of the  weakly focussed  laser wave.  The experiment at CALA  aims to prepare a monoenergetic bunch by the end of the acceleration process and this will occur if  the electron beam is short enough to experience an approximately uniform accelerating field. This in turn takes place at distances smaller than  the  plasma wavelength $\lambda_{\mathrm{p}}=2\pi \omega_{\mathrm{p}}^{-1}$ with $\omega_{\mathrm{p}}=( e^2 \mathpzc{n}_e/m)^{1/2}$ referring to the plasma frequency. Indeed,  in  the blow-out regime of the LWFA  $\mathpzc{n}_b>\mathpzc{n}_e$, with $\mathpzc{n}_b$ refering to  the electron beam density,    both  the axial $\sigma_{z}$ and radial $\sigma_{r}$ extensions  of the  witness bunch are bound by  $\sigma_{z,r}<\lambda_{\mathrm{p}}/(2\pi)$   \cite{Esarey1,Esarey2,Lobet}. Under such a circumstance,  the beam spreading  characterized by a broadening angle $\theta_{e^-}$  has been estimated to be  of the order of $\theta_{e^-}\approx0.5$ mrad, while the distance travelled  by the electrons  towards the high-Z target will be  set to $L_{e^-}=10$ cm. 

 Simulations carried out  in Ref.~\cite{Salgado}  provide evidences that only $\sim 1\%$ of the electrons in the beam  will produce  bremsstrahlung  radiation.  Upon penetrating a high-Z target with  thickness  $L_{\mathrm{T}}$ much smaller than the characteristic radiation length $L_{\mathrm{rad}}$ of the material [$L_{\mathrm{T}}\ll L_{\mathrm{rad}}$], the emission of  bremsstrahlung photons  takes  place within the electron beam  volume which   undergoes almost no spatial spreading  due to the  ultra-relativistic nature of its constituents. Hence, we will assume that the longitudinal extension of the bremsstrahlung burst  amounts to $\sigma_z$ as well.    In this context, the spreading angle of  generated radiation  can be approximated by the inverse electron Lorentz factor $\theta_{\gamma} \approx1/\gamma_e= m/E_0\sim \mathcal{O}(1)$ mrad.  As a consequence  the vast majority of  the bremsstrahlung  photons are emitted tangentially  to the direction of propagation of the incident electron beam. The energy spectrum of bremsstrahlung photons, which are produced  by electrons passing through a  solid high-Z target within the thin target  approximation  reads  \cite{PartPhysGroup,Tsai}
\begin{equation}\label{thin}
I_\gamma(f,\ell)=E_0\frac{d\mathcal{N}_\gamma}{d\omega^\prime} \approx \frac{\ell}{f}(4/3-4f/3+f^2),
\end{equation} where $\mathcal{N}_\gamma$ is the number of emitted photons by a radiating electron,   $f = \omega'/E_0$ refers to the  normalized photon  energy   of the emitted $\gamma$ radiation and $\ell = L_{\mathrm{T}}/L_{\mathrm{rad}}$ denotes the normalized target thickness. In our numerical studies we shall assume a target made of tungsten for which the radiation length is $L_{\rm rad}=3.5$ mm. 

Unless stated otherwise, we use Eq.~\eqref{thin} throughout this study.
\begin{figure}[ht]
\includegraphics[width=0.4\textwidth]{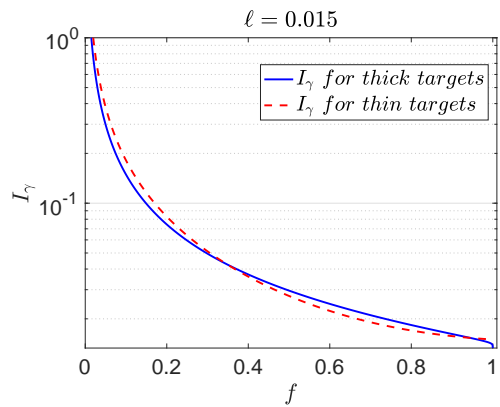}
\caption{\label{fig0} Bremsstrahlung spectra according to Eqs.~\eqref{thin} (red dashed) and \eqref{thick} (blue solid).}
\end{figure}
However, we should mention at this point that another approximative formula for the bremsstrahlung spectrum is available in the literature \cite{Tsai} which, in contrast to Eq.~\eqref{thin}, is well suited for thicker targets with $\ell<2$: 
\begin{equation}\label{thick}
I_\gamma(f,\ell)\approx \frac{(1-f)^{4\ell/3}-\mathrm{e}^{-7\ell/9}}{f(7/9 +4/3 \mathrm{log}(1-f))}.
\end{equation}
We compare the outcomes from Eqs.~\eqref{thin} and \eqref{thick} for $L_\mathrm{T}=50\rm \mu m$ ($\ell=0.015$)  in Fig.~\ref{fig0}. Here, the dashed red line accounts for the thin target approximation, whereas  the blue solid holds for thicker targets. It can be seen that deviations appear mainly in the region $f\leq 0.2$, where both curves show an unrealistic divergence in the bremsstrahlung distribution function. Hence, the impact of this low energetic part of the spectrum on the pair production yield must be kept negligible. Furthermore, the thin target approximation does not manifest the characteristic steep decrease at the point $f\approx 1$ stemming from the fact that no photons can be emitted with energy greater than the energy of the incident electrons. Also this region of the spectrum therefore needs to be considered with care. Consequently, the contributions of incident electrons with different energies to the pair creation rate  by considering separately  thin and tick targets are studied  at the beginning of Sec.~\ref{Results}.

As described above, in the second stage the  bremsstrahlung $\gamma$ burst  collides with a high-intensity laser pulse of optical frequency $\omega \sim \mathcal{O}(1)$ eV and large value of $\xi\gg 1$. The collision will take place at some distance $L \sim \mathcal{O}(1)$ m from the target under a collision angle $\phi$ between the bremsstrahlung beam and the strong laser pulse.  Observe that, due to the spreading, the bremsstrahlung radiation covers  a  volume (see  Fig.~\ref{fig1}):
\begin{equation}\label{vgamma}
V_{\gamma} \approx \pi  \sigma_z \bar{r}^2\quad\mathrm{with}\quad
\bar{r}=\frac{r_{\mathrm{min}}+r_{\mathrm{max}}}{2}
\end{equation} 
the average radius of the truncated cone formed by the bremsstrahlung burst. Here, the maximal and minimal  radii  are $r_{\mathrm{max}}\approx\theta_{\mathrm{rms}}\sigma _z+r_{\mathrm{min}}$  and  $r_{\mathrm{min}}=\sigma_r+L_{e^-}\theta_{e^-}+L\theta_{\mathrm{rms}}$, respectively  with $\sigma_r<\lambda_{\mathrm{p}}/(2\pi)$ denoting the transversal extension of  the witness  beam and $\theta_{\mathrm{rms}}=(\theta_{e^-}^2+\theta_{\gamma}^2)^{\nicefrac{1}{2}}$ counting for the root-mean squared of the spreading angle. We note that, in the planned experiment at CALA, $\bar{r}\approx r_{\mathrm{min}}\approx L_{e^-}\theta_{e^-}+L\theta_{\mathrm{rms}}\approx 320\;\mu\rm m$ (see Table I).

\subsection{Laser field profiles}\label{laser}

The  element  of the  setup  that  is left  to be described is  the high-intensity laser field which takes part  in the second stage of the experiment.  We will   suppose this pulse background  propagating  along the $z-$axis with a  linear polarization  characterized by the vector  $\pmb{\epsilon}=(1,0,0)$.  Consequently, the nontrivial electric field component  to be specified is $E_x=B_y$. While in the course of our calculations  various strong field profiles are analyzed,  the field shape resulting from  the  paraxial approximation  is adopted as a reference model \cite{Salamin}:  
\begin{equation}\label{fieldparaxial}
E_x = \mathcal{E}_0\frac{\mathrm{e}^{-\left(\sqrt{2\ln(2)}\frac{(t-z)}{\tau}\right)^2}}{\sqrt{1+\zeta(z)^2}}
 \mathrm{e}^{-\left(\frac{r}{w(z)}\right)^2}\mathrm{sin}(\Phi).
\end{equation} 
Here, $w(z)=w_0\sqrt{1+\zeta(z)^2}$ stands for the beam width which depends on the longitudinal coordinate $z$ via the factor $\zeta(z)=z/z_R$. In this context, $z_R=\pi w_0^2/\lambda$ is the Rayleigh length, whereas  $w_0$ refers to the beam waist size at the focal point ($z=0$).  The non-trivial Gaussian  dependence on $r^2=x^2+y^2$  accounts for the  transversal  behaviour of the pulse, whereas its temporal extension $\tau$ is taken at FWHM from the intensity.  Moreover, the  pulse phase is  
\begin{equation}\label{Phi}
\Phi=\omega(t-z)-\zeta(z)\frac{r^2}{w^2(z)}+\mathrm{arctan}(\zeta).
\end{equation}

The beam  energy carried by this pulse, which in practice is fixed and does not change by focusing  into various field profiles,  can be calculated from the associated local power: 
\begin{multline}
P(t,z)=\frac{\mathcal{E}_0^2}{2}\frac{\pi w_0^2}{2}  \mathrm{e}^{-2\left(\sqrt{2\ln(2)}\frac{\varphi}{\omega\tau}\right)^2}    \\
 \times \Big\{1 -\frac{1}{1+\zeta(z)^2}\left[\cos(2\varphi)-\zeta(z)\sin(2\varphi)\right]\Big\}
\end{multline}
with $\varphi=\omega(t-z)$. 
Once the integration over time  is carried out,  one obtains the pulse energy
\begin{equation}\label{Wn2G}
W_{\mathrm{G}} \approx \frac{\mathcal{E}_0^2}{2} \frac{\pi w_0^2}{2} \frac{\tau}{2}\sqrt{\frac{\pi}{\ln(2)}}
\end{equation}
with the accuracy up to a term decreasing exponentially in $(\omega\tau)^2$,  provided  $\omega\tau\gg1$. The expression above will be used  to  adjust the laser intensity and focussing parameters  linked to other field models while keeping their energies  equal to Eq.~\eqref{Wn2G}.

In addition to the previous focused Gaussian pulse, we shall adopt a description for  the  strong laser field which relies on  a  pulsed plane wave  model. The field associated with this scenario can be read off from Eq.~\eqref{fieldparaxial}  when the limit  $w_0\to \infty$ is taken. Explicitly, 
  \begin{equation}\label{fieldPulse}
  E_x(\varphi) =  \mathcal{E}_0\mathrm{e}^{-\left(\sqrt{2\ln(2)}\frac{\varphi}{\omega\tau}\right)^2}\sin\left(\varphi\right):= \mathcal{E}_0\psi(\varphi).
\end{equation} 
In contrast to the paraxial approximation a plane wave is infinitely  extended perpendicularly to the direction of propagation and as a consequence,  the associated  beam  power would formally diverge. In order to perform a fair comparison,  the  infinite transversal beam area can be parametrized conveniently by  $A_{\mathrm{int}}$. 
 In such a scenario the instantaneous beam power results $P(\varphi)=E^2_x(\varphi) A_{\mathrm{int}}$ and  the corresponding  energy carried by the beam reads
\begin{equation}\label{Wn2}
W_{\mathrm{pw}} \approx \frac{\mathcal{E}_0^2}{2} A_{\mathrm{int}} \frac{\tau}{2}\sqrt{\frac{\pi}{\ln(2)}}.
\end{equation}
A comparison with Eq.~\eqref{Wn2G} allows us to identify  the effective interacting area  $A_{\mathrm{int}}=\pi w_0^2/2$.  Implicitly, this means that the strong field is truncated transversally, i.e.  $E_x(\varphi) \to E_x(\varphi) \Theta(r)\Theta(w_0/\sqrt{2}-r)$ with $\Theta(x)$ denoting the unit step function.  Observe that in the limit $\omega \tau \to \infty$ the plane wave in Eq.~\eqref{fieldPulse} becomes monochromatic with a power $P(\varphi)=\mathcal{E}_0^2\sin^2(\varphi) A_{\mathrm{int}}$.  In this context, the beam energy equals $W_{\mathrm{mpw}}=\langle P\rangle T_{\mathrm{int}}$, where $\langle P\rangle = IA_{\mathrm{int}}$
is its mean power with $I = \mathcal{E}_0^2/2$ referring to the time-averaged intensity. Upon a comparison between $W_{\mathrm{mpw}}$ and Eq.~\eqref{Wn2}  we end up with $T_{\mathrm{int}} = \frac{\tau}{2}\sqrt{\frac{\pi}{\ln(2)}}$. 

 \begin{figure}[ht]
\includegraphics[width=0.45\textwidth]{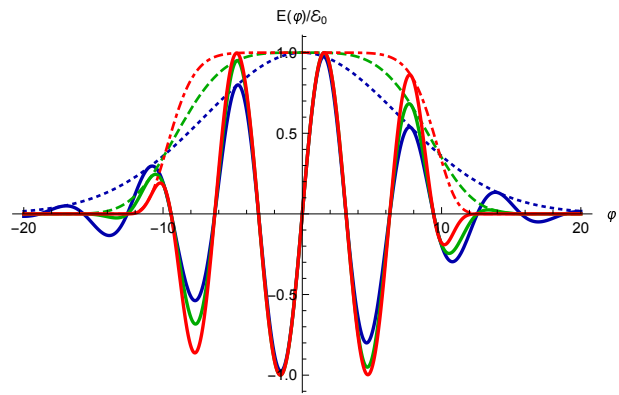}
\caption{\label{GaussianTime} Plane-wave  pulses ($w_0 \to \infty$) with Gaussian  and super-Gaussian-profiles   with $n=4$ and $n=8$ are shown in blue,  green and red, respectively. While the pure Gaussian envelope is dotted, the corresponding  modulated functions linked to  $n=4$  and $n=8$  are  dashed  and  dot-dashed. These envelopes are given as references. This picture has been generated by setting  $\tau=5\; \rm fs$  and $\omega =1.55\; \rm eV $ .}
\end{figure}

The procedure is extendable to paraxial pulses modulated by super-Gaussian profiles, which are characterized by  higher powers within the time-dependent exponent of Eq.~\eqref{fieldPulse}, i.e. when $2\to n$ with $n= 4, 8, \ldots$ (see Fig.~\ref{GaussianTime}). Observe that, as $n$ increases, the field profiles linked to super-Gaussian models  acquire  plateaus which enable to reach the peak intensity  several times as compared  to the case modulated by the standard Gaussian function.  Likewise, the aforementioned growing of $n$ reduces gradually the ramping (deramping) interval, making  its slope steeper than in the Gaussian model. The corresponding energies for $n= 4$ and $n= 8$  are
\begin{equation}\label{wSGTime}
\begin{split}
&W_{n=4} \approx \frac{\mathcal{E}_{0}^2}{2} \frac{\pi w_0^2}{2} \frac{\tau2^{1/4}\Gamma\left(\frac{5}{4}\right)}{\sqrt{\ln(2)}}, \\
& W_{n=8} \approx \frac{\mathcal{E}_{0}^2}{2}  \frac{\pi w_0^2}{2}  \frac{\tau2^{3/8}\Gamma\left(\frac{9}{8}\right)}{\sqrt{\ln(2)}}
\end{split}
\end{equation}
with $\Gamma(x)$ denoting the Gamma function \cite{NIST}. Observe that as long as we refer to a common laser system the peak intensity will vary when the pulse energy is kept fixed.

Super-Gaussian profiles can also be used when  modeling the transverse shape of the wave.  However, in contrast to the previous scenario,  the intensity  linked  to these beams  at  the focal plane $z=0$ is \cite{SuperGaussian, SGFG}
\begin{equation}
I(r)=I_0\mathrm{e}^{-2\left(\frac{r}{w_0}\right)^m}
\end{equation}
with $m\geq 2$ and peak intensity $I_0=\mathcal{E}_0^2$.  Correspondingly,  a pulse transversally focussed by  a super-Gaussian profile has a non-vanishing electromagnetic field component  of the form
\begin{equation}\label{SuperGaussianTrans}
E_x = 
\mathcal{E}_0\mathrm{e}^{-\left(\sqrt{2\ln(2)}\frac{\varphi}{\omega\tau}\right)^2}
 \mathrm{e}^{-\left(\frac{r}{w_0}\right)^m}\sin\left(\varphi\right).
\end{equation}  
For $m=2$, the formula above describes the leading order term of the paraxial field (see Eq.~\eqref{fieldparaxial})
in $z/z_R\ll 1$, which is a good approximation for the focal inner region. 
In the limit of $m\to \infty$ the transversal part tends to a rectangular function. 
The pulse energy for $m=4$ and $m=8$ results into
\begin{equation}\label{WnSG4}
\begin{split}
&W_{m=4} \approx \frac{\mathcal{E}_0^2}{2} \frac{\pi^{3/2} w_0^2}{2^{3/2}} \frac{\tau}{2}\sqrt{\frac{\pi}{\ln(2)}}, \\
& W_{m=8} \approx \frac{\mathcal{E}_0^2}{2} \frac{\pi w_0^2\Gamma\left(\frac{5}{4}\right)}{2^{1/4}} \frac{\tau}{2}\sqrt{\frac{\pi}{\ln(2)}} .
 \end{split}
\end{equation}
We remark that an analytical expression for the transversal super-Gaussian beams does not exist outside the focal plane $z=0$. Hence, Eq.~\eqref{SuperGaussianTrans} can only be applied when longitudinal focusing of the laser beam can be disregarded. 

\subsection{Pair creation by bremsstrahlung photons in a strong laser pulse  within the  locally constant field approximation} \label{NCPs}

We shall suppose that the production process is dominated by the space-time region in which the strong field condition $\xi(\pmb{x},t)\gg1$ holds.  Under such  circumstances  the characteristic pair formation length   $l\sim \lambda/(\xi\pi)$ turns out to be much smaller than $\lambda$,  enabling  an  effective  description  in which the laser background can be treated locally  as   a constant field whose electric and magnetic field strengths are orthogonal and equal.  As a consequence,  the local probability rate per unit  of volume  for producing  a  pair  by a single bremsstrahlung $\gamma$ photon can be approximated by \cite{Bulanov, ReportFedotov} 
\begin{equation}
\label{LCFA}
\begin{split}
\left.\frac{d\mathpzc{P}}{dt dV}\right \vert_{\xi(\pmb{x},t)\gg1}\approx\left.R(\kappa)\right\vert_{\xi\to\xi(\pmb{x},t)},
\end{split}
\end{equation}  where $R(\kappa)$ is  the transition rate per volume of the pair production process  in a constant crossed field \cite{NikishovRitus,RitusReview,NikishovRitus2}:  
\begin{equation}
\label{R}
R(\kappa) = -\frac{\alpha m^2}{6\sqrt{\pi}\omega' V_\gamma} \int_{1} ^\infty \frac{du (8u +1)}{u\sqrt{u(u-1)}} \frac{\Phi'(z)}{z}.
\end{equation} Here  $\kappa = kk'\xi/m^2$ denotes the quantum non-linearity  parameter  with  $k'=(\omega', \pmb{k}^\prime)$ and $k=(\omega, \pmb{k})$  referring to  the  corresponding  four-momentum of the  $\gamma$ quanta  and the laser wave.  Observe that the local density  rate depends on the derivative $\Phi'(z)=d\Phi/dz$ of  the Airy function $\Phi(z) = \frac{1}{\sqrt{\pi}}\int_0^\infty dt \cos(\frac{t^3}{3}+zt)$ with argument  $z=(4u/\kappa )^{2/3}$. We remark that the expression above applies as long as $\xi \gg \mathrm{max}\{1,\kappa^{1/3}\}$ holds and perturbation theory is still valid, i.e. if the condition  $\alpha\kappa^{2/3}<1$ is fulfilled  with $\alpha\approx1/137$ referring to the fine structure constant \cite{NikishovRitus3,Narozhny,Baumann,Mironov,Podszus,Ilderton}.

In the following we shall assume that  the mean collision angle between the  bremsstrahlung  burst and  strong laser pulse is $\phi=\pi$. As we are interested in comparing the pair yields stemming from  various external field models, we shall discuss in this section particularities of  the corresponding expressions for the number of produced pairs per radiating  electron:  
\begin{equation}\label{NGaussian}
N \approx\int_0^{E_0}d\omega^\prime\langle\mathpzc{P}\rangle\frac{d\mathcal{N}_\gamma}{d\omega^\prime}=\int_0^{1}df\langle\mathpzc{P}\rangle I_{\gamma},
\end{equation}  
where  $I_\gamma$ are  given  in  Eqs.~\eqref{thin}  and  \eqref{thick} and the change of variable $\omega'=fE_0$ has been carried out. In the formula above  $\langle\mathpzc{P}\rangle=\frac{1}{2\delta\phi}\int_{\phi-\delta\phi}^{\phi+\delta\phi}d\tilde{\phi}\;\mathpzc{P}(\tilde{\phi})$ is the  pair production probability    (see Eq.~\eqref{LCFA}):
\begin{equation}\label{LCFAI}
\mathpzc{P}(\phi)\approx\int_{\Gamma}dtdV\left.R(\kappa)\right\vert_{\xi\to\xi(\pmb{x},t)}
\end{equation} 
averaged over the collision angle. Hereafter we suppose  a deviation  $\delta \phi$ due to  the spreading of  both the decaying  gamma quantum   and the laser pulse   much  smaller than $\phi$, i.e.  $ \phi\gg\vert\delta\phi\vert$, in which case  the approximation $\langle\mathpzc{P}\rangle\approx \mathpzc{P}$ applies.

\begin{figure}[ht]
\begin{flushleft}
\includegraphics[width=0.58\textwidth]{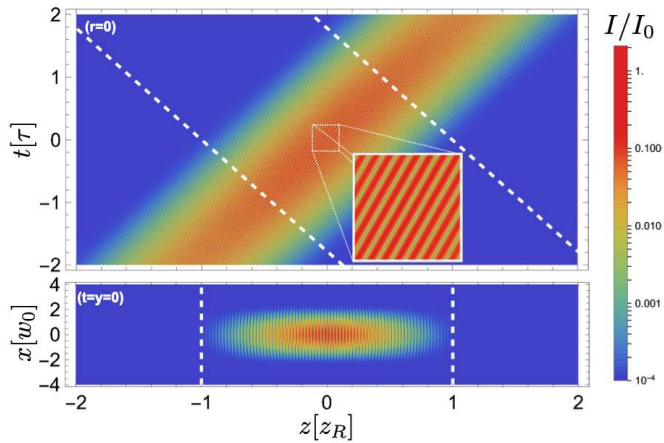}
\end{flushleft}
\caption{\label{regionintegration} 
Behavior of the laser intensity  $I(\pmb{x},t)=E^2(\pmb{x},t)$  of a paraxial pulse  (see Eq.~\eqref{fieldparaxial})  in  $t$ and $z$ (upper panel) and in $x$ and $z$ (lower panel). Here, the dashed  lines represent wavefronts of bremsstrahlung radiation separated between each other by a distance  $\sigma_z =2z_R$.  The inset  of the upper panel  reveals the oscillatory feature of the strong field along  $t$ and $z$ axes.  We have used the same benchmark values and notation as in Table I.  For these values and $I_0 \approx 2 \times10^{22} \ \mathrm{W/cm^2}$, the strong field condition $\xi(\pmb{x},t)\gg1$  translates into $I(\pmb{x},t)/I_0\gg 2 \times 10^{-4}$ and $z_R = 15.7 \  \mu \mathrm{m}$. }
\end{figure}

Noteworthy,  the domain of integration in Eq.~\eqref{LCFAI}, i.e. $\Gamma$,  is defined by  the interaction region where  the strong field condition $\xi(\pmb{x},t)\gg1$ is fulfilled. Observe that  for an optical laser with frequency $\omega=1.55\; \rm eV$ and peak intensity $I_0 \approx 2 \times10^{22} \ \mathrm{W/cm^2}$  (i.e. $\xi=70$), this translates into $I(\pmb{x},t)/I_0\gg 2\times 10^{-4}$.  As  Fig.~\ref{regionintegration} exhibits,  this spacetime  sector  is characterized by  oscillations of the strong field  modulated by both the  focussing and pulse shape functions. Despite  the cumbersome  form of $\Gamma$, the fast damping of $R(\kappa)$  for $\kappa\ll1$ allows us to extend  this domain  to  the whole  interaction  region, where the bremsstrahlung beam and laser pulse overlap, without introducing an appreciable error. Indeed, in the presence of  a strong pulse,  the integration region turns out to be determined  by  the most separated wavefronts of the bremsstrahlung burst. This requires  that in our model the radiation mode with frequency $\omega'$ extends over the phase interval  $-\frac{1}{2}\omega^\prime\sigma_z \leq k^\prime x\leq\frac{1}{2}\omega^\prime\sigma_z$.  In Fig.~\ref{regionintegration},  this restriction translates into  a band   encompassed  between the dashed  lines: $z_\pm=-t\pm\frac{1}{2}\sigma_z$.  At this point we should stress that the derivation of  Eq.~\eqref{R}  relies  on a monochromatic plane-wave  wavefunction for the decaying   $\gamma$ quantum. This means that our model assumes the axial $\sigma_z$ and radial $\bar{r}$  extensions of the bremsstrahlung radiation to substantially exceed the associated wavelength $\lambda^\prime=2\pi \omega^{\prime-1}$, which is safely fulfilled.

 With all these details to our disposal we find that  the scenario where  the strong  laser field  turns out to be described by a paraxial Gaussian pulse (see Sec.~\ref{laser}), the probability  can be  expressed as
\begin{equation}
\begin{split}
\mathpzc{P}\approx2\pi\int_{-\infty}^{\infty}dt\int_0^{\infty}r dr\int_{-t-\frac{1}{2}\sigma_z}^{-t+\frac{1}{2}\sigma_z}dz\; \left.R(\kappa)\right\vert_{\xi\to\xi(\pmb{x},t)},\label{apP}
\end{split}
\end{equation} 
where cylindrical coordinates have been adopted. The combination of  Eqs.~\eqref{NGaussian}-\eqref{apP}  with \eqref{thin}  included,   constitutes the starting point of our numerical analysis.  We note  that beyond the paraxial approximation, the cylindrical symmetry of the laser pulse is broken through nontrivial dependences on the azimuthal angle (see Eqs.~\eqref{BPField} and \eqref{BPFieldB}). In such a case,  the factor $2\pi$ in the expression above has to be replaced by an integration over the aforementioned angle.

Next, we consider  models  in which the strong field background depends only on the phase $\varphi$ (see Eq.~\eqref{fieldPulse} and the discussion that follows it).  This scenario  can be formulated conveniently via light-cone coordinates:  $x_\pm =\frac{1}{\sqrt{2}}(t\pm z)$, $\pmb{x}_\perp=(x,y)$  \cite{Neville,Mitter}. As a consequence,  the strong field phase becomes $\varphi=k_+x_-$ with $k_+=(k^0+k^3)/\sqrt{2}=\sqrt{2}k_0$. In this context,  $kk^\prime=k_+k_-^\prime$, whereas the phase of the gamma quantum $k^\prime x=k^\prime_-x_+$.  We note that this set of variables  allows  us  to express the pair production probability as
\begin{equation}
\begin{split}
&\mathpzc{P}=\frac{\omega^\prime}{k_+k_-^\prime}V_{\mathrm{int}}\int_{-\infty}^{\infty} d\varphi\left.R (\kappa)\right\vert_{\xi\to\xi(\varphi)}\\
&\quad=-\frac{\alpha m^2}{6\sqrt{\pi}kk^\prime}\frac{V_{\mathrm{int}}}{V_{\gamma}}\int_{-\infty}^{\infty} d\varphi\int_1^\infty \frac{du(8u+1)}{u\sqrt{u(u-1)}}\frac{\Phi^\prime(z)}{z}.
\end{split}\label{Npw}
\end{equation} 
Notice that  the establishment of  the second line required the substitution of   Eq.~\eqref{R} explicitly. As before, $z=(4u/\kappa(\varphi))^{2/3}$ with $\kappa(\varphi)=\kappa \vert \psi (\varphi)\vert$. In this expression  $V_{\mathrm{int}}$  is   the interacting volume,  the precise form of which depends on the strong field model.\footnote{For  a  collision geometry other than  head-on, $V_{\mathrm{int}}$ can depend on the crossing angle $\phi$.} Indeed, if the latter turns out to be a plane wave,  $V_{\mathrm{int}}$ and $V_\gamma$ coincide and the resulting expression agrees with the outcome resulting from  Eq.~($33$)  of Ref.~\cite{Meuren}. However, in contrast to our procedure, the expression in the aforementioned reference  was obtained  from the imaginary part of the vacuum polarization tensor in a plane-wave background via the optical theorem.  Now, if the strong pulse is truncated transversally with a size $w_0/ \sqrt{2} < \bar{r}$, the interaction  volume  turns out to be  determined by the region  occupied by the external field. In this case  $V_{\mathrm{int}}=A_{\mathrm{int}}\sigma_z$ with  $A_{\mathrm{int}}$  given  below Eq.~\eqref{Wn2},  and $V_{\mathrm{int}}/V_\gamma\approx w_0^2/(2 \bar{r}^2)$, where the result given below Eq.~\eqref{vgamma} has been used. The ratio between $V_{\mathrm{int}}$  and $V_\gamma$  accounts for the fraction of bremsstrahlung photons that interact with the strong  laser pulse and for the parameters assumed in Sec.~\ref{Results} amounts to $\approx 2 \times 10^{-5}$. 

Lastly, if the  external  field is approximated by  $E_x(\varphi)=\mathcal{E}_0\Theta(\frac{1}{2}\Delta\varphi-\varphi)\Theta(\varphi+\frac{1}{2}\Delta\varphi)$, the background turns out to be a constant crossed field. Under such condition the pair creation probability from Eq.~\eqref{Npw} reads
 \begin{equation}\label{rateCCF}
 \mathpzc{P} = T_{\mathrm{int}}V_{\mathrm{int}} R(\kappa)
 \end{equation}
provided the relation $T_{\mathrm{int}}= \frac{\tau}{2}\sqrt{\frac{\pi}{\ln(2)}}=\Delta\varphi/(2\omega)$ holds (see below Eq.~\eqref{Wn2} for $T_{\mathrm{int}}$).

Now, the experiment  put  forward by  CALA aims to probe the  nonperturbative strong-field regime  ($\xi\gg1,\;\kappa\sim 1$)  of the  Breit-Wheeler process.  In order to elucidate the behaviour of $R$ in this limit, we first exploit the relation $\Phi'(z) =-\frac{z}{\sqrt{3\pi}}K_{2/3}\left(\frac{2}{3}z^{3/2}\right)$ with $K_\nu(x)$ denoting the modified Bessel function of the second kind \cite{NIST}.  As a consequence, Eq.~\eqref{R} can be written as 
\begin{equation}\label{ratenum}
\begin{split}
R = \frac{\alpha m^2}{6\pi \omega' V_\gamma} \frac{8}{3^{3/2}\kappa}\int_{\frac{8}{3\kappa}} ^\infty dp \frac{3\kappa p +1}{p^{3/2} \sqrt{p-\frac{8}{3\kappa}}} K_{2/3}(p),
\end{split}
\end{equation}
where the change of variables  $p=8u/(3\kappa)$ has been carried out. For $\kappa \sim 1$ the main contribution in the integral results from the region $p\sim 8/(3\kappa)$. By approximating  the integrand  with its most slowly decreasing part, we end up with
\begin{equation}\label{rateApprox}
\begin{split}
&R_{\kappa \approx 1} 
\approx \frac{\alpha m^2}{6\pi\omega' V_\gamma} \sqrt{8\kappa}\int_{\frac{8}{3\kappa}} ^\infty dp \frac{K_{2/3}(p)}{ \sqrt{p-\frac{8}{3\kappa}}}  \\
&\quad= \frac{\alpha m^2}{\pi\omega'V_\gamma}\left(\frac{2}{3}\right)^{3/2} K_{7/12}\left(\frac{4}{3\kappa} \right)K_{1/12}\left(\frac{4}{3\kappa} \right).
\end{split}
\end{equation} 
\begin{figure}[ht]
\includegraphics[width=0.5\textwidth]{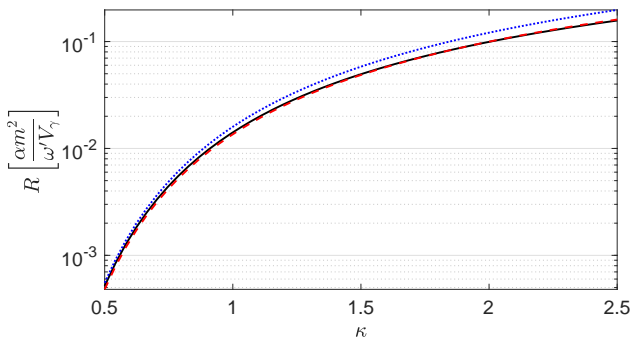}
\caption{\label{fig2} Comparison of the numerically evaluated rate as given in Eq.~\eqref{R} (black solid) with the analytical $\kappa \approx 1$ asymptote from  Eq.~\eqref{rateApprox} (red dashed) and the limiting case $R_{\kappa \ll 1} \approx \frac{\alpha m^2}{8\omega'V_{\gamma}}\left(\frac{3}{2}\right)^{3/2}\kappa \mathrm{e}^{-\frac{8}{3\kappa}}$  that is valid for $\kappa\ll 1$ (blue dotted) \cite{NikishovRitus2}.}
\end{figure}
In Fig.~\ref{fig2} we show the behaviour of the rate given in Eq.~\eqref{ratenum} as a function of $\kappa$ (black solid). For comparison, we have added the rate linked to Eq.~\eqref{rateApprox} in red dashed style and the one corresponding to the case $\kappa \ll 1$ in blue dotted. It is worth  remarking  that in the region of $\kappa \in [1.5,2.5]$ the error introduced by Eq.~\eqref{rateApprox} lies below $2\%$ and grows to approximately $10\%$ for $\kappa \ll 1$.  This analysis reveals that $R_{\kappa \approx 1}$ provides a good description of the pair production rate for the present study and will be adopted in the forthcoming  numerical calculations.


\section{Results and discussion}\label{Results}

\subsection{Comparison of different field models\label{CDFM}}

In this section we use the expressions derived so far  to provide estimates for the pair yield by an incident radiating bremsstrahlung electron assuming various laser field models. We shall consider an ideal collision characterized by  a perfect synchronization. To avoid longitudinal mismatching,  the extension of the  bremsstrahlung burst  will be  chosen so that,  at $t=0$,  it covers fully  the  strong field region of the laser pulse.  An examination of Fig.~\ref{regionintegration}  indicates that the conservative value of  $\sigma_z=2z_R\approx 31.4\;\mu\rm m$ --- which is  taken hereafter as a reference parameter --- guarantees  the previous condition. Observe that   the  upper bound  discussed at the beginning  of Sec.~\ref{bspe}, i.e. $\sigma_z<\lambda_{\mathrm{p}}/(2\pi)$ \cite{Esarey1,Esarey2},  implies  that the  plasma density has to satisfy the condition   $\mathpzc{n}_e<2.25\times 10^{16}\; \rm cm^{-3}$  for  consigning  a  monoenergetic witness bunch. We note that this  limitation  remains within the ballpark   $\mathpzc{n}_e<\mathpzc{n}_b$ with $\mathpzc{n}_b\approx10^{18}\; \rm cm^{-3}$   established  in Ref.~\cite{Salgado}. Unless explicitly stated otherwise, we use the benchmark parameters  listed  in Table~I, and a counterpropagating geometry ($\phi=\pi$). 

\begin{table}[h]\label{table1}
\begin{tabularx}{8.3cm}{|X|c|}
\hline
\hline
Incident electron energy $E_0$ & $ 2.5\;\rm  GeV$ \\
\hline
Distance travelled by the bunch  $L_{e^-}$ & 0.1 m \\
\hline
Incident electrons collimation angle $\theta_{e^-}$ & 0.5 mrad \\
\hline
Normalised target thickness $\ell$ & 0.015 \\
\hline
Distance travelled by bremsstrahlung $L$ & 0.5 m \\
\hline
Wavelength  of the strong pulse $\lambda$ & 0.8 $\mu$m \\
\hline
Pulse waist size $w_0$ & 2 $\mu$m \\
\hline
Pulse length  $\tau$ & 30 fs \\
\hline
Laser  intensity parameter $\xi$ & 70\\
\hline
Laser repetition rate &  0.1 Hz \\
\hline\hline

\end{tabularx}
\caption{ The benchmark parameters envisaged at the experiment to be carried out  at CALA  in Ref.~\cite{Salgado}. These values are  adopted hereafter.}
\end{table}

\begin{figure}[ht!]
\includegraphics[width=0.5\textwidth]{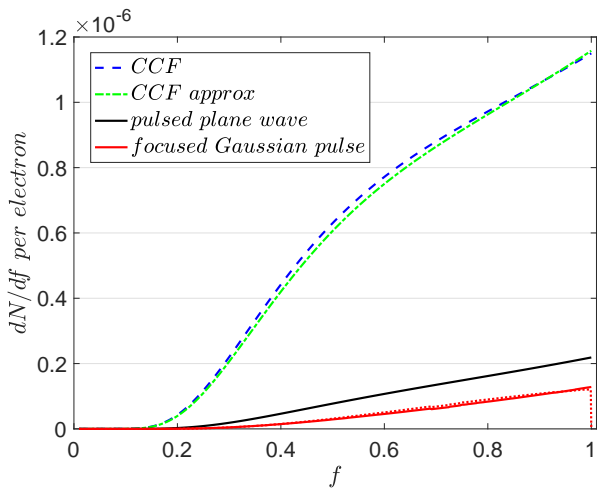}
\caption{\label{fig4} Differential number of pairs in dependence on the scaled energy of bremsstrahlung photons for $\xi=70$.  We use the same benchmark values and notation as in Table  I.}
\end{figure}

We begin our study by analysing the dependence of the created particle distribution on the  bremsstrahlung photon energy. This behaviour is summarised in Fig.~\ref{fig4}. Depending on the underlying field description, the curves (blue dashed for constant crossed fields, black solid for pulsed plane wave and red solid for pulsed Gaussian) show  the corresponding number of  produced pairs.  The figure was generated in the following way: for the blue curve we consider only a fraction of the ---infinitely extended--- field enclosed in the spacetime volume $V_{\mathrm{int}}T_{\mathrm{int}}$, which contains the same energy as the Gaussian laser pulse (see Eq.~\eqref{rateCCF}). In the case of the pulsed plane wave we proceed analogously and restrict the spatial components to $V_{\mathrm{int}}$. While the constant crossed field description provides the most optimistic prediction, the incorporation of a finite laser duration and a laser focusing diminishes the expected yield by about a factor 5 and 10, correspondingly. 

Additionally, all curves in Fig.~\ref{fig4} increase as the energy of bremsstrahlung photons grows meaning that higher energetic photons facilitate the studied pair production process. 
For the sake of completeness we note that at the right edge of the spectrum the curves should have a maximum at  $f \approx 1$ and fall sharply to zero afterwards. This trend is closely linked to the fact that no $\gamma$ photons can be produced with energy exceeding $E_0$ and, accordingly, no pairs can be created. The absence of the falling is a result of using the thin target approximation Eq.~\eqref{thin}  in $N$ (see Refs.~\cite{PartPhysGroup} and \cite{Blackburn}). For comparison, Fig.~\ref{fig4} includes a red dotted curve, which results when the thick target approximation in Eq.~\eqref{thick} is applied  and a focused Gaussian pulse model adopted. As both red curves  lie very close to each other and the contribution of the low energy range $f \in [0,0.2]$ is negligible we conclude that in the regime of interest, the approximation in Eq.~\eqref{thin} is well applicable and, thus, will be used  throughout this study (see discussion below Eq.~\eqref{thick}). In Fig.~\ref{fig4} we have also contrasted  the differential number of pairs  resulting from the expressions in Eqs.~\eqref{R} (blue) and \eqref{rateApprox} (green dotted). The close overlapping of the curves  supports the applicability of the asymptotic formula $R_{\kappa\approx 1}$  in the regime of interest.

\begin{figure}[ht]
\includegraphics[width=0.5\textwidth]{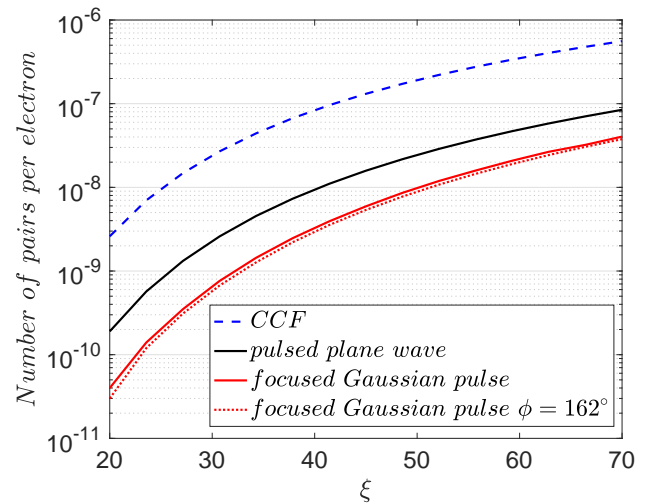}
\caption{\label{fig5} Pair yield per radiating electron for constant crossed fields (blue dashed), pulsed plane wave (black) and pulsed Gaussian profile (red). The same benchmark parameters and notation of Table  I have been used.}
\end{figure}

The behaviour of the expected total number of created pairs as a function of $\xi$ is depicted in Fig.~\ref{fig5}. Results stemming from Eq.~\eqref{rateCCF}, with the inclusion of \eqref{apP} and \eqref{Npw} are shown in dashed blue, solid black and solid red, respectively. This figure includes a red dotted curve which has been obtained by setting the collision angle $\phi = 9\pi/10$, as it is planned in the experiment described in Ref.~\cite{Salgado}. The corresponding rate has been obtained by performing a rotation of the integration region depicted in Fig.~\ref{regionintegration}. In line, the limitation in the phase of the bremsstrahlung radiation translates into the following limits in $z$: $z_\pm = -(t \mp \frac{1}{2}\sigma_z)/\vert \cos(\phi)\vert +r\vert\tan(\phi)\vert$.  Observe that the last term may be neglected as long as the longitudinal extension of bremsstrahlung beam is larger than the laser pulse length. We remark that the outlined procedure represents a good approximation whenever the collision geometry is close to the  counterpropagating case.

It can be seen that the incorporation of more realistic field configurations, as it was indicated previously in Fig.~\ref{fig4}, modifies the expected outcome by lowering the pair production yield. We remark that the reduction effect originates solely from the laser field description as energy is kept constant for all field shapes (see Sec.~\ref{laser}). While in the case of the constant crossed field the intensity is kept high and constant within the whole interaction spacetime volume (which can be read off from Eq.~\eqref{Wn2G}), the field intensity linked to a Gaussian profile changes from its maximum at the center of the interaction volume to minimal values at its edges. Hence, regardless the consideration of the whole spacetime in the integration of Eq.~\eqref{LCFAI}, the intensity gradient has a significant impact on the pair creation yield. The particles are mainly produced in the rather small high-field region close to the focal point -- whereas the extended outer regions of the pulse, where the pair production is negligible, still contribute to the pulse energy.

Let us particularise the analysis for $\xi=70$ which corresponds to an intensity of $I \approx 10^{22} \ \mathrm{W/cm^2}$. Under such circumstances the expected number of created positrons  per incident radiating electron in a single laser shot is $4 \times 10^{-8}$. With experimental techniques available nowadays such as laser wakefield acceleration, electron bunches with up to $\approx 1$ nC charge can be generated \cite{Leemanns2019, Karsch}. For an envisaged energy of $E_0$ = 2.5 GeV, we expect bunches of several pC \cite{Salgado}. Therefore, with a $10$ pC electron bunch $\approx 0.03$ pairs per laser shot can be observed, if we assume that $1 \%$ of incident electrons will emit a bremsstrahlung photon (see discussion in Sec.~\ref{bspe}). Hence, when taking into account a laser repetition rate of $0.1$ Hz a yield of 10 Breit-Wheeler pairs is expected per hour. This outcome turns out to be much smaller then a prediction reported in Ref.~\cite{Blackburn}, where for $\xi = 30$ up to $10^4$ pairs per laser shot and a pC of 2 GeV incident electrons were estimated. However, in contrast to our scenario the prediction given in \cite{Blackburn} assumes that the divergence of the bremsstrahlung ray can be counteracted by focusing the incident electron bunch.  In our case, the pair yield is reduced by a factor $V_{\mathrm{int}} / V_{\gamma} \sim 10^{-5}$ that results from the beam divergences.

\begin{figure}[ht]
\includegraphics[width=0.5\textwidth]{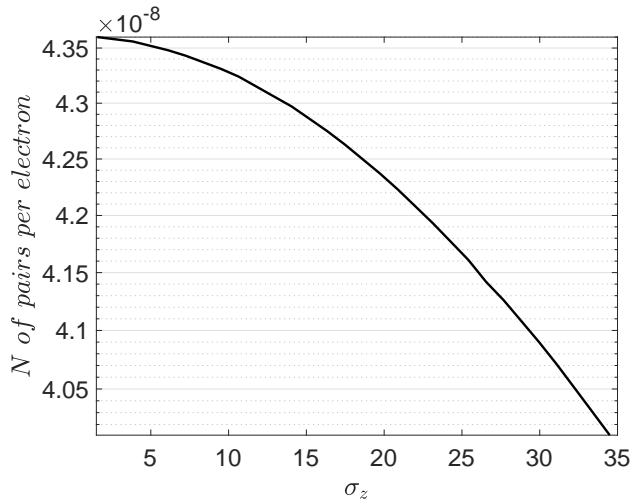}
\caption{\label{BremsThickness} Dependence of the pair yield on the thickness of the  bremsstrahlung bunch $\sigma_z$ for parameters given in Table I. Moreover, the number of gamma photons in the pulse is kept constant.}
\end{figure}
As a next step we analyse, in Fig.~\ref{BremsThickness}, the dependence of the pair yield on the thickness of the bremsstrahlung burst. For obtaining the depicted results we modelled the laser as a paraxial Gaussian pulse (see Eq.~\eqref{fieldparaxial}). The curve exhibited in this picture shows a downward tendency, which is caused by the longitudinal focusing. Observe that, when the bremsstrahlung bunch is shorter than the laser focal region $2z_R$, more $\gamma$ photons experience high intensity of the pulse, provided a good synchronisation is achieved. On the contrary, for longer bremsstrahlung bunches the contributions from lower intensity regions will decrease the pair yield at the edges of the interaction volume. Notice that a factor of $\sigma_z$ in $V_\gamma$ (see Eqs.~\eqref{R} and \eqref{vgamma}) counteracts the naive expectation that the number of created pairs will always grow with increasing $\sigma_z$.
\subsection{Contributions from different focal regions \label{CFDFR}}

\begin{figure}[ht!]
\includegraphics[width=0.5\textwidth]{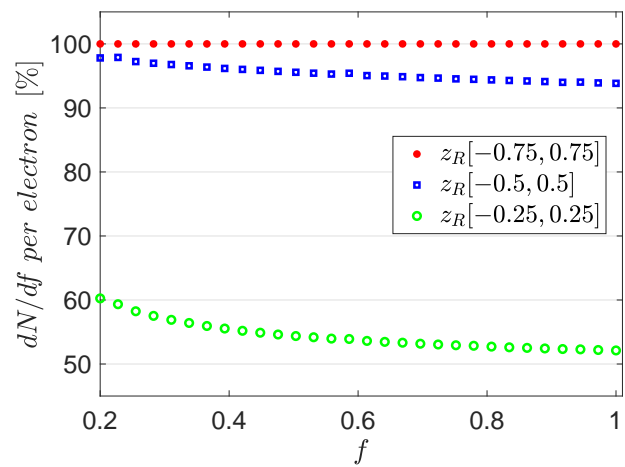}
\caption{\label{fig4a} Differential number of pairs for different branches of focal regions for a focused Gaussian pulse for $\xi=70$. The other parameters are given in Tab.~I.}
\end{figure}

Next, we examine the contributions to the number of pairs stemming from different focal regions. The outcome of this investigation is exhibited in Fig.~\ref{fig4a}.  These curves have been obtained when modelling the laser field as a paraxial Gaussian pulse and changing the integration limits in Eq.~\eqref{apP} to $z\in [-0.75,0,75]z_R$, $z\in [-0.5,0,5]z_R$, $z\in [-0.25,0,25]z_R$, respectively, and $t_{\pm} = -z \pm \frac{1}{2}\sigma_z$.  Here, the patterns in red filled circles, blue open boxes and green open circles manifest the fraction of the pair yield stemming from the reduced integration regions to $z \in$ $[-0.75,0.75]z_R$, $[-0.5,0.5]z_R$ and $[-0.25,0.25]z_R$, respectively. Our assessment reveals that the relative contribution from the innermost region is the higher, the lower the bremsstrahlung photon energy is. This is understandable because, for rather low-energy $\gamma$ photons, a large value of the laser field is very crucial to yield a sizable pair production signal. However, the major contribution to the total number of pairs stems from high bremsstrahlung energies (see Fig.~\ref{fig4}), where the particles are created in the following proportions: while $52\%$ result from $|z|\le 0.25 z_R$, the doubled region with $|z|\le 0.5z_R$ gives $94\%$, and practically $100\%$ of the pair production is contained in $\vert z\vert\le 0.75z_R$.

\begin{figure}[ht!]
\includegraphics[width=0.5\textwidth]{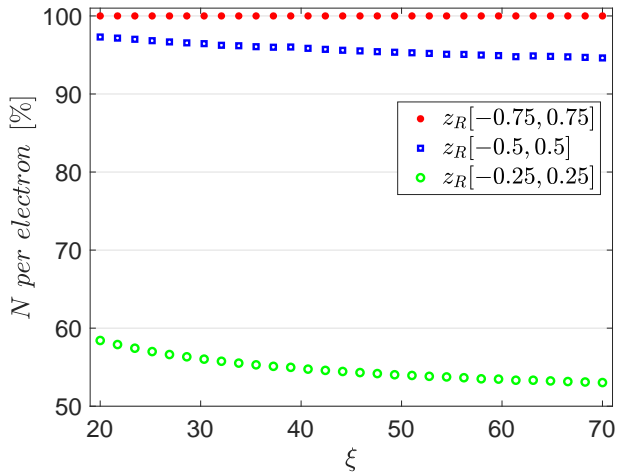}
\caption{\label{fig4b} Percentage of created particles from different focal regions for a focused Gaussian pulse. The other parameters are given in Tab.~I.}
\end{figure}

Fig.~\ref{fig4b} shows how the percentage of the particles produced in different focal regions varies with the laser intensity parameter. Here, we model the laser pulse and change the integration limits as described previously for Fig.~\ref{fig4a}. The trend exhibited by the curves indicates that with the growing of $\xi$ the importance of the outer zones increases as the high intensity region that facilitates the pair production is extended to the whole Rayleigh length. To be more precise, while the innermost region (green open circles) accounts for about $53 \%$ of created pairs for $\xi=70$, its impact increases to about $58 \%$ when the intensity parameter is lowered to $20$. This tendency results from the fact that, for rather low $\xi$, the local quantum nonlinearity parameter $\kappa$ reaches significant values (close to 1), as required for a sizeable pair production signal, only in the inner focal region. Outside this region, $\kappa$ quickly falls far below 1 and the pair production is suppressed, accordingly. In contrast, when $\xi$ is large, the local value of $\kappa$ reaches a sizeable level over a more widespread region where the pair production can occur with significant probability. (Note that the slope of the curves in Fig.~\ref{fig5} decreases with increasing $\xi$, so that local changes of the field strength in a Gaussian pulse become less crucial when $\xi$ is large.)

\begin{figure}[ht!]
\includegraphics[width=0.5\textwidth]{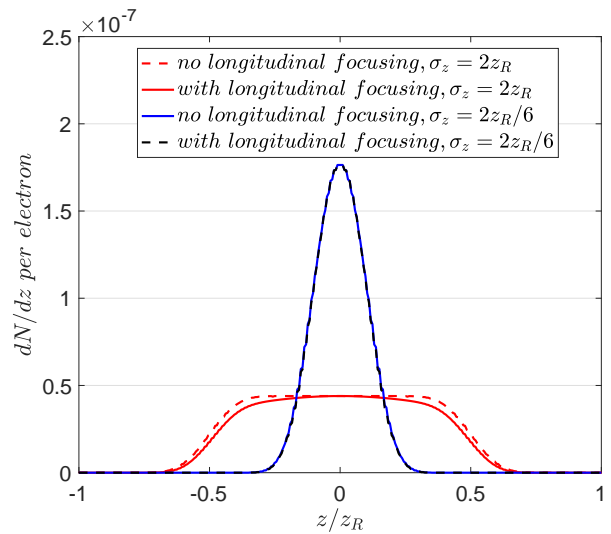}
\caption{\label{fig4bPrime} Distribution of created pairs in the longitudinal direction.  We use the same benchmark values and notation as in Table  I.  }
\end{figure}

A comparison of the distribution of created pairs along the $z-$axis  with $\sigma_z = 2z_R$ and $\sigma_z = 2z_R/6$ is shown in Fig.~\ref{fig4bPrime}. In both cases the dashed curves ignore the longitudinal focusing in the description of the laser field, which is achieved by omitting dependences on $\zeta(z)$ in the paraxial field model in Eq.~\eqref{fieldparaxial}. Conversely, the solid curves incorporate this effect. Also here, the  integration in $t$ was limited by $t_\pm = -z \pm \frac{1}{2}\sigma_z$.
Observe that the red curves deviate from each other outside the zone $z\in [-0.25,0.25]z_R$. The outcomes for a shorter bremsstrahlung bunch with $\sigma_z = 2z_R/6$ is shown in blue and black. Here, the effect of the longitudinal focusing is absent as the interaction occurs in the innermost focal region (read the discussion below Fig.~\ref{BremsThickness}). Moreover, the larger maximum of the blue curve as compared to the red one at $z=0$ can be understood  as a direct  consequence of the shorter bremsstrahlung extension: the  number of $\gamma$  quanta  that experiences the region of highest field strength turns out to be larger.

\subsection{Focusing effects \label{FEEFFF}}

\begin{figure}[ht!]
\includegraphics[width=0.5\textwidth]{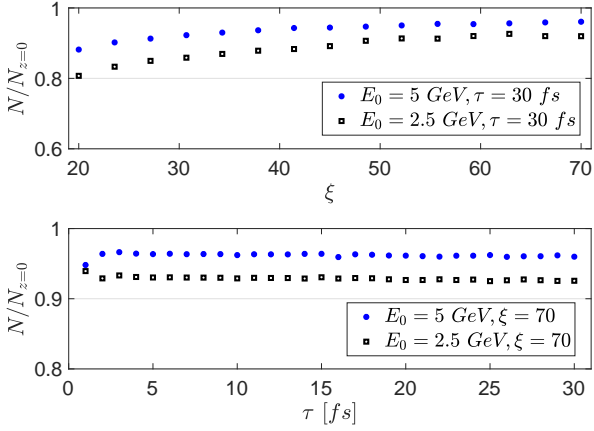}
\caption{\label{fig4bPrimePrime} Dependence of the ratio between the number of produced pairs with ($N$) and without ($N_{z=0}$) longitudinal focusing on the intensity parameter $\xi$ (upper panel) and the pulse length $\tau$ (lower panel). The other parameters are given in Tab.~I.}
\end{figure}

The significance of a wider longitudinal focusing  is studied further in Fig.~\ref{fig4bPrimePrime}, where  the  ratio between the number of produced pairs  linked to  models  with ($N$) and without ($N_{z=0}$) longitudinal focusing   is  exhibited as a function of the  intensity parameter (upper panel) and the pulse duration (lower panel).   In both panels the  black and blue dotted curves correspond to  $E_0= 2.5$ GeV and $E_0=5 $ GeV. However, while  the upper panel  has been obtained by setting the laser pulse duration to  $\tau=30 \ \mathrm{fs}$,   the lower panel follows by setting $\xi=70$.

On the one hand,  the upper panel manifests that the increasing energy of the incident electrons allows us to neglect the longitudinal focusing: for $\xi=70$ the relative error $1-N/N_{z=0}$ drops from $\sim 8 \%$ to $\sim 4 \%$. This tendency, similarly to the effect caused by the intensity, originates in the dependence of  $\kappa$ on the considered parameters (see i.e. Eq.~\eqref{kappapw} with $\omega'=fE_0$). When the values of $E_0$ and $\xi$ are so that $\kappa \approx 1$, the corresponding rate as given in Eq.~\eqref{rateApprox} has a slope less pronounced  than in  the case where $\kappa \ll 1$  (see also Fig.~\ref{fig2}). On the other hand, the curves in the lower panel of Fig.~\ref{fig4bPrimePrime} show  a  plateau for $\tau \gtrsim 3$ fs. 

\begin{figure}[ht]
\includegraphics[width=0.5\textwidth]{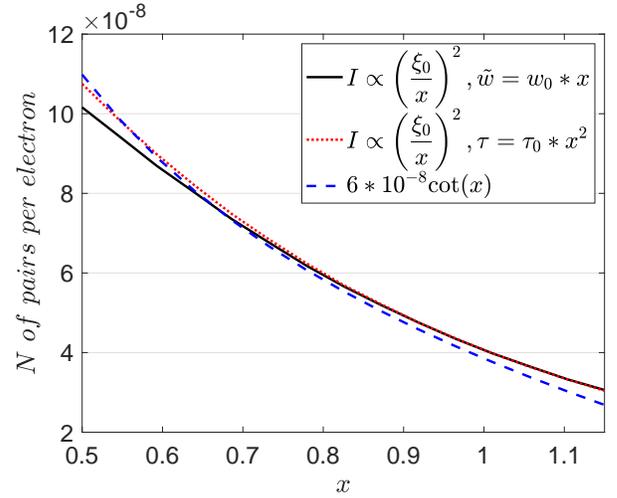}
\caption{\label{fig6} Impact of focusing when keeping the laser pulse energy constant.
Here, the pair yield is maximal for all curves at the smallest considered value $x = 0.5$, which corresponds to the largest intensity parameter $\xi=2\xi_0$, minimal beam waist $\tilde{w}=w_0/2$ (black) and  minimal pulse duration $\tau=\tau_0/4$ (red dotted). The comparison is made for the reference values $\xi_0=70$, $w_0 = 2 \ \mu$m and $\tau_0=30$ fs corresponding to $x=1$.}
\end{figure}

Next, we examine the implication of the transversal laser focusing  in more detail. According to Eq.~\eqref{Wn2G}, the laser energy $W_{\mathrm{G}}\propto Iw_0^2$, where $I=\mathcal{E}_0^2/2$ stands for the average laser intensity. Hence, we can study the impact  on the number of pairs yielded   by  varying  simultaneously  both the laser intensity $I$ and the beam waist  $w_0$ while keeping $W_{\mathrm{G}}$ constant as achieving higher intensity demands stronger focusing, i.e. narrower beam. 
The outcome can be seen through the blue curve in Fig.~\ref{fig6} for $w_0 \in  [1,2.3] \ \mu\mathrm{m}$ and $I \in [0.86, 4.2]\times 10^{22} \ \mathrm{W}/\mathrm{cm}^2$. Here, the smallest value of the beam waist corresponds to the highest intensity. The graph shows a pronounced  declining pattern  as the intensity decreases gradually while the waist size grows simultaneously.  This can be explained because $N\propto \cot (x)$, where $x$ is the proportionality factor that  diminishes $\xi$  and  increases $w_0$. The introduction of this parameter encodes two paths to optimise the production of pairs: either by increasing the intensity or by increasing the interaction volume. However, as it can be seen from Fig.~\ref{fig6}, the former benefits the process more than the latter in the considered parameter range.
Thus, an optimisation of the volume quotient $V_{\mathrm{int}}/V_{\gamma}$ should be achieved by, for example, collimating the incident electron beam with a quadrupole magnet (decreasing $\theta_{e^-}$) or via a faster deflection of electrons which have passed the high-Z target (decreasing $L$) and not by loose focusing. Observe that the red dotted curve in Fig.~\ref{fig6} depicts the dependence of the yielded pair number on the variations of intensity and pulse duration when keeping the pulse energy constant $W_{\mathrm{G}}\propto I\tau^2$. In analogy to the effect of the beam waist, we see that the consideration of longer pulses at the cost of smaller intensity does not benefit the production of pairs. 
\begin{figure}[ht]
\includegraphics[width=0.5\textwidth]{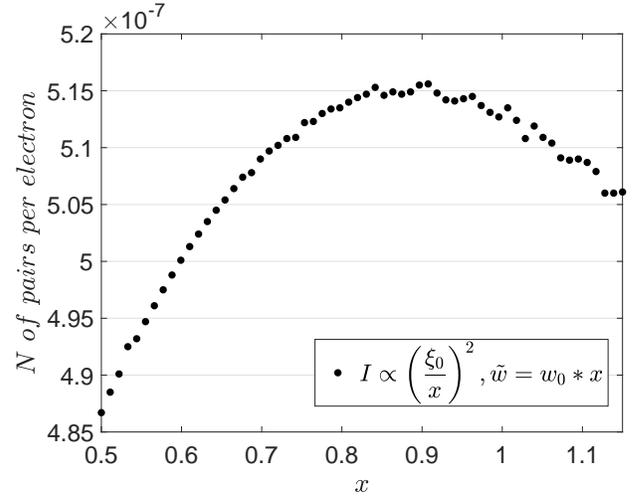}
\caption{\label{fig6a} Number of produced pairs as a function of the laser intensity and the beam waist while keeping the pulse energy constant. The optimal intensity point is found at $x \approx 0.9$,  corresponding to $I\approx 3\times10^{22}\ \mathrm{W/cm^2}$ ($\xi=120$) at $\tilde{w}=0.9\ \mathrm{\mu m}$. The comparison is made for the reference values $\xi_0=110$, $w_0 = 1 \ \mu$m corresponding to $x=1$ and $E_0=10$ GeV.}
\end{figure}

By applying a similar procedure as described above, an optimal intensity for the considered setup is found when the  incident electron energy is chosen as $E_0 = 10$ GeV. Fig.~\ref{fig6a} exhibits the maximum at $x\approx0.9$, which corresponds to $\xi\approx120$ ($I\approx 3\times10^{22}\ \mathrm{W/cm^2}$) and $\tilde{w}\approx0.9 \ \mathrm{\mu m}$. It is worth noting that for $x\lesssim 0.9$, although the strong laser field is more tightly focused, its increased intensity does not guarantee a maximisation of the pair production yield.  On the contrary, the respective decrease in the interaction volume outweighs the effect from the increased intensity and results in a shrinking number of created pairs. This study extends the outcome found in Ref.~\cite{Salgado}, where a similar analysis was carried out for pair production by a monoenergetic $\gamma$ beam and an intense, not focussed laser pulse.  As a consequence, the optimal intensity established there was around $I\approx 10^{22}\ \mathrm{W/cm^2}$ for $\omega'=2.5$ GeV and $\tilde{w}=2 \ \mathrm{\mu m}$. 

\begin{figure}[ht]
\includegraphics[width=0.5\textwidth]{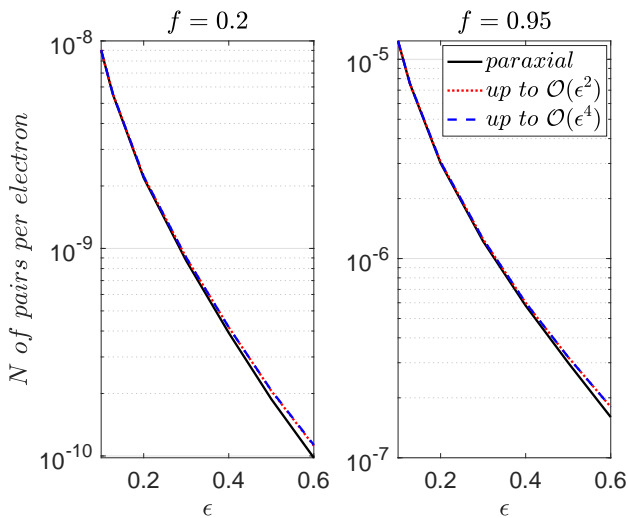}
\caption{\label{fig7} Deviation in the number of created pairs when considering Gaussian pulse in paraxial approximation (black solid)  and beyond paraxial pulses (dotted red and dashed blue) for different diffraction angles $\epsilon$ and $\xi=70$. Here, the energy of incident photons is fixed to $\omega'=500$ MeV (left panel) and $\omega'=2.357$ GeV (right panel).}
\end{figure}

As it was pointed out previously, the results so far were generated within the paraxial approximation, which is valid as long as the diffraction angle $\epsilon=2/(w_0\omega) \ll 1$. 
However, at the points $w_0=1 \ \mu$m and $w_0=0.5 \ \mu$m for $\lambda=0.8 \ \mu$m  (see Figs.~\ref{fig6} and \ref{fig6a}) we reach $\epsilon\approx 0.255$ and $\epsilon\approx 0.51$ correspondingly, which brings the scenario closer to the diffraction limit. Hence, in order to evaluate the extent to which our calculations are well suited, higher order terms in $\epsilon$ have been incorporated (see App.~\ref{BeyondPA}). The outcome of this study is summarized in Fig.~\ref{fig7}. In this picture, we see the number of created pairs for two particular energies of the  bremsstrahlung spectrum, i.e. we do not average over it but rather keep the photon energy constant at $500$ MeV (left panel, $f=0.2$) and $2375$ MeV (right panel, $f=0.95$).  In both cases small deviations in the number of created pairs per bremsstrahlung photon start to appear for $\epsilon \gtrsim 0.4$. This fact agrees with the extent of modifications that appear in the electric and magnetic fields at $\epsilon \gtrsim 0.5$ when going beyond the paraxial approximation in Ref.~\cite{Salamin}, where these fields were initially introduced. Hence, the paraxial approximation is well applicable for the laser parameters envisaged in the present study.

\begin{figure}[ht]
\includegraphics[width=0.5\textwidth]{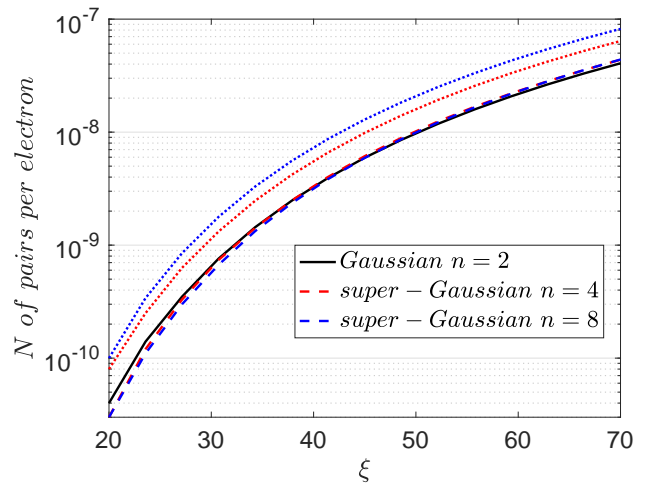}
\caption{\label{fig8} Number of produced pairs for laser pulses with super-Gaussian time envelopes (red, blue dashed) compared to standard Gaussian with intensity parameter $\xi=70$ (black solid) when keeping the laser energy at the standard Gaussian level. Dotted curves result when the value of $\xi$ is kept the same for all pulse shapes.}
\end{figure}

\subsection{Super-Gaussian profiles \label{SGPE}}

We wish to determine the extent to which the results discussed so far are sensitive to the chosen Gaussian time envelope. To evaluate the deviations, we incorporate in the paraxial model (as given in Eq.~\eqref{fieldparaxial}) a super-Gaussian time profile (see Sec.\ref{laser} and Fig.~\ref{GaussianTime}). These envelopes are characterized by steeper edges and broader plateau regions as contrasted to a standard Gaussian. In order to make a fair comparison we keep the energy of super-Gaussian pulses (exemplarily $n=4$ and $n=8$) equal to the standard paraxial pulse. This has been achieved by adjusting the intensities to 
\begin{equation}
\begin{split}
& I_{n=4}=I\frac{\sqrt{\pi}}{2^{5/4}\Gamma\left(5/4\right)} \approx 0.82 I, \\
& I_{n=8}=I\frac{\sqrt{\pi}}{2^{11/8}\Gamma\left(9/8\right)} \approx 0.73 I,
\end{split}
\end{equation}
 where $I$ stands for the standard paraxial intensity. Thus, we check a standard Gaussian at a particular $\xi$ against super-Gaussian time envelopes with lower effective intensity parameters (as $\xi = \frac{\vert e\vert}{m\omega}\sqrt{2I}$). 
 The outcome is depicted in Fig.~\ref{fig8}, where the $\xi$ parameter of the standard Gaussian  has been varied. Here, the number of created pairs stemming from the latter time profile (solid black line) is compared to super-Gaussians with $n=4$ (red dashed) and $n=8$ (blue dashed). While the three curves show an upward trend, the super-Gaussians overpass the pure Gaussian result at $\xi \approx 40$. This agrees with our previous finding that, for relatively lower values of $\xi$, the pair production is optimized by the highest field intensity, whereas for large $\xi$ values, the process may benefit more strongly from a broadening of the high-field interaction zone. Moreover, the dotted lines in Fig.~\ref{fig8} result from the study, where no modification of the intensity was undertaken. As expected, in this case the super-Gaussians lead to a higher pair yield in the whole intensity range, with the outcome for $n=8$ exceeding the one for $n=4$. 

\begin{figure}[ht]
\includegraphics[width=0.5\textwidth]{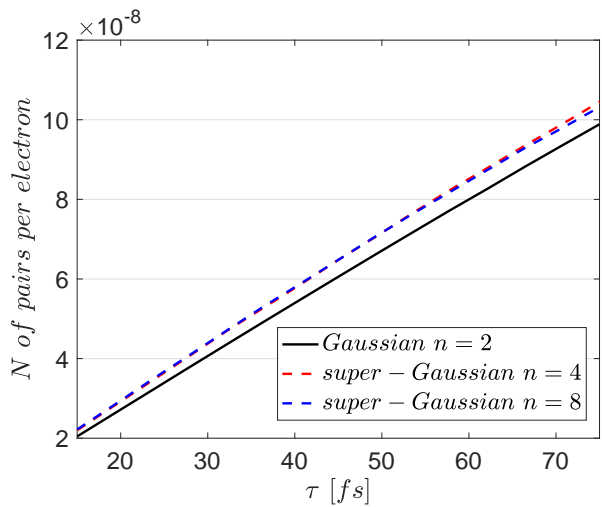}
\caption{\label{fig9} Number of produced pairs for laser pulses with super-Gaussian  time envelopes (red, blue dashed) compared to standard Gaussian (black solid) when keeping the laser energy at the standard Gaussian level. Hence, the intensity parameters read $\xi=70$ for $n=2$, $\xi=63.5$ for $n=4$ and $\xi=59.6$ for $n=8$.}
\end{figure}

Further details are presented in Fig.~\ref{fig9}, where the dependence of the pair yield on the pulse duration is shown. Here, as it was done previously the laser pulse energy is kept constant while the field shape is varied by taking a Gaussian with $n=2, \xi=70$ (black solid) and super-Gaussians with $n=4, \xi =63.5$ (red dashed) and $n=8, \xi=59.6$ (blue dashed). While the number of created particles grows with increasing pulse duration for every investigated temporal profile, the broader envelopes lead to a higher pair production yield. Hence, for the considered values of $\xi$, the effect of  increasing the effective interaction time $T_{\mathrm{int}}$ over the plateau region (see Eq.~\eqref{wSGTime}) outweighs the decrease in intensity, which was needed to keep the pulse energies equal.

\begin{figure}[ht]
\includegraphics[width=0.5\textwidth]{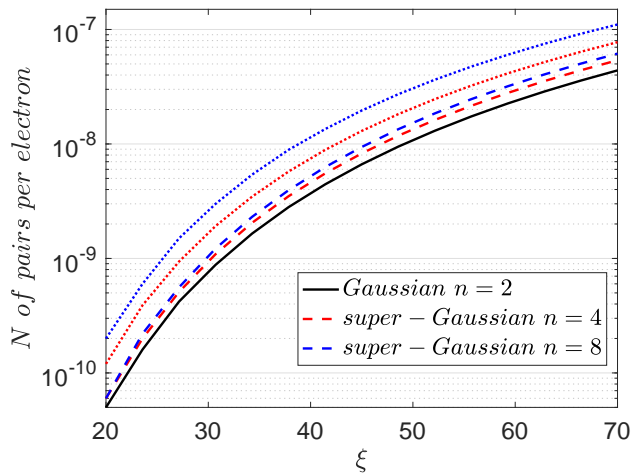}
\caption{\label{fig10} Number of produced pairs for laser pulses with super-Gaussian spatial envelopes (red, blue dashed) in the transverse plane compared to standard Gaussian with intensity parameter $\xi$ (black solid) when keeping the laser energy at the standard Gaussian level . Dotted curves result when the value of $\xi$ is kept the same for all pulse shapes.}
\end{figure}
Now, we proceed to study the impact of changing interaction volume when considering  super-Gaussian profiles in the transversal focusing. To assess  the extent to which the number of produced pairs is modified  due to this feature, we adjust   their  intensities so that the energy carried by these fields  coincides with that linked to the standard Gaussian in paraxial approximation. Hence,  the corresponding laser intensity parameters  turn out to be 
\begin{equation}
\begin{split}
&\xi_{n=4}=\xi (2/\pi)^{1/4}\approx 0.89\xi ,\\
&\xi_{n=8}=\xi (2^{3/4}\Gamma\left( 5/4\right))^{-1/2} \approx 0.81\xi.
\end{split}
\end{equation}
At this point, we have varied  the $\xi-$parameter associated with the paraxial model to evaluate  the number of yielded pairs. We note that the analytical expression for the super-Gaussians applies only in the $z=0$ plane, which means that the consequences linked to the longitudinal focusing are ignored. As it has been indicated previously (see Fig.~\eqref{fig4bPrime}), for parameters used in the present study this assumption will not lead to a substantial error.
The results of this evaluation are  summarized in Fig.~\ref{fig10}. Here, the outcome related to a super-Gaussians with  $n=4$   is exhibited in red dashed, whereas the one linked to $n=8$ appears in blue dashed style. Observe that both curves lie  above the black solid curve corresponding to the paraxial result. The effect shown in Fig.~\ref{fig10}, although small, provides some hints regarding the importance of optimising the interaction area: the effect of decreasing the intensity from $I = 10^{22} \ \mathrm{W/cm^2}$ to $I_{n=4} = 8.4\times10^{21} \ \mathrm{W/cm^2}$ and $I_{n=8} = 6.9\times 10^{22} \ \mathrm{W/cm^2}$ may be outweighed by broadening the laser beam. 
Additionally, Fig.~\ref{fig10} shows  dotted curves  which  describe the number of pairs yielded when the intensity is kept constant for for all pulse models. As before, under this condition, the super-Gaussians give larger outcomes than the standard Gaussian throughout, with the pair yield being largest for the super-Gaussian  with $n=8$.

\section{Conclusion\label{Conclusions}}

Summarizing, we have investigated  how the non-linear Breit-Wheeler pair  creation  process in the nonperturbative regime with $\xi \gg 1$ depends on the model adopted for describing the strong field of the laser.  Our analysis  has  been  focused  on a  setup   which combines   highly energetic  $\gamma$ photons produced from  bremsstrahlung  and  a high intensity  laser  pulse. We have shown that,  in such a scenario,   an optimization  of the yield closely depends on both the laser intensity  and the extension  of the interaction region.

Throughout the paper we have contrasted the outcomes resulting from  different laser field models, including   the constant crossed field, the  plane-wave  and  the   paraxial Gaussian pulse. This analysis indicates that  transversal and longitudinal  focusing  of the beam are  important features  to be  taken  into  account to quantitatively describe upcoming  experiments characterized by  $\xi\gg 1$ and $\kappa\approx 1$. In order to gain clarity of their role, the percentage of pairs produced in various  focal zones along the longitudinal direction has been elucidated. We have shown that, for relatively moderate values of $\xi$, the majority of particles is produced in the innermost focal region, whereas the contributing interaction zone grows when $\xi$ is increased. Besides, the consequences of broadening  the transversal beam profile and  the pulse length were investigated separately by adopting  super-Gaussian models for the strong  laser field.   However, this study  has revealed no significant difference with respect to the paraxial scenario.

Moreover, the influence of changes in the laser intensity, pulse duration and energy of the incident bremsstrahlung electrons has been considered and the crucial importance of an optimized overlap between the transverse extent of the bremsstrahlung beam and the laser beam waist was emphasized. For the parameters of an envisaged future experiment, relying on incident electrons of $E_0= 2.5$ GeV energy to generate the bremsstrahlung and laser pulses of $800$ nm wavelength, $\xi = 70$ and $30$ fs pulse duration \cite{Salgado}, we expect the creation of about 0.03 pairs per  $10$ pC of incident charge and laser shot. This number appears to be resolvable with the advanced detection technologies available nowadays.

\begin{acknowledgments}
This work has been funded by the Deutsche Forschungsgemeinschaft (DFG) under Grant No. 416699545 within the Research Unit FOR 2783/1. The authors thank F. Salgado, K. Grafenstein, D. Seipt, F. Karbstein, J. Farmer and  A. Pukhov  for useful discussions.
\end{acknowledgments}

\appendix

\section{Time and space dependent quantum non-linearity parameter}\label{LocalKappa}

In order to take into account the structure of the strong field we express the quantum non-linearity parameter $\kappa = |e|\sqrt{-(F_{\mu\nu}k'^{\nu})^2}/m^3$ in terms of the electromagnetic field tensor $F_{\mu\nu}$. As a consequence 
\begin{equation}
\begin{split}\label{kappa}
\kappa&= \frac{|e|\omega'\mathcal{E}_0}{m^3} \left[\left(\frac{B_z}{\mathcal{E}_0} \sin(\phi) +\frac{B_y}{\mathcal{E}_0} \cos(\phi)\right)^2 +\left(\frac{E_x}{\mathcal{E}_0}\right)^2 \right. \\
 & + \left(\frac{E_y}{\mathcal{E}_0}\right)^2 + \left(\frac{E_z}{\mathcal{E}_0}\right)^2 -\left(\frac{E_y}{\mathcal{E}_0} \sin(\phi) +\frac{E_z}{\mathcal{E}_0} \cos(\phi)\right)^2\\
&+  2\frac{B_x}{\mathcal{E}_0}\left( \frac{E_y}{\mathcal{E}_0}\cos(\phi) - \frac{E_z}{\mathcal{E}_0} \sin(\phi)\right) +\left(\frac{B_x}{\mathcal{E}_0}\right)^2 \\
&\left. + 2 \frac{E_x}{\mathcal{E}_0}\left( \frac{B_z}{\mathcal{E}_0} \sin(\phi) -\frac{B_y}{\mathcal{E}_0} \cos(\phi)\right) \right]^{1/2},
\end{split}
\end{equation}
where $E_i$ and $B_i$ are the electric and magnetic field components, respectively (see Eq.~\eqref{fieldparaxial}). Here,  $\phi$ denotes the collision angle, which refers to the z-axis: we assume the geometry, in which the strong laser pulse propagates with the wave vector $\pmb{k}=\omega\pmb{e}_z$ (see Fig.~\ref{fig1}).

\begin{figure}[ht]
\includegraphics[width=0.5\textwidth]{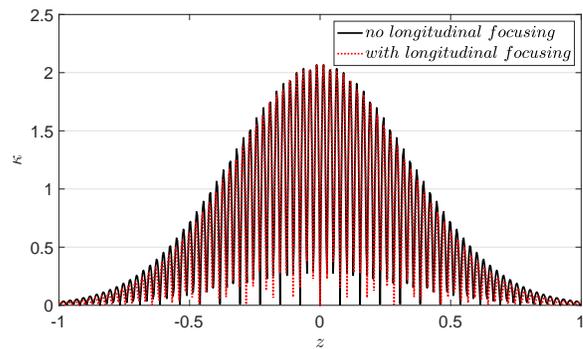}
\caption{\label{fig11} Local values of the quantum non-linearity parameter $\kappa$ for Gaussian pulses with (red dotted) and without (black solid) longitudinal focusing at $t=0$, $r=0$.}
\end{figure}

When the expression above is particularised to the case described in Sec.~\ref{laser}, we end up with
\begin{equation}\label{kappapulse}
\kappa= \frac{|e|\omega'}{m^3} [1-\cos(\phi)]|E_x(t,r,z)|.
\end{equation}

In Fig.~\ref{fig11} the dependence of $\kappa$ on the longitudinal coordinate $z$ is displayed. This picture has been generated by taking into account Eq.~\eqref{fieldparaxial} with $r=0$ and $t=0$. Hence, in the parameter range discussed in this work the local values of $\kappa$ encompass the interval between $0$ and roughly $2$. In the limiting case of $w_0\to\infty$ Eq.~\eqref{kappapulse} reproduces the formula for the quantum non-linear parameter in a plane wave background $E_x(t,z)=E_x(\varphi)=E_0\psi\left( \varphi\right)$:
\begin{equation}\label{kappapw}
\kappa_{\mathrm{pw}}(\varphi) = \kappa|\psi(\varphi)|,
\end{equation}
where $\kappa=\omega \omega'[1-\mathrm{cos}(\phi)]\xi/m^2$ refers to the standard quantum non-linearity parameter that arises in calculations dealing with a monochromatic plane wave \cite{Meuren} and the function $\psi(\varphi)$ is defined in Eq.~\eqref{fieldPulse}.

\section{Gaussian pulse beyond paraxial approximation}\label{BeyondPA}

The  electric and magnetic fields of a strong Gaussian  pulse are modified by higher order contributions in the diffraction angle $\epsilon = w_0/z_R$. According to Ref.~\cite{Salamin}, up to fourth order they read
\begin{widetext}
\begin{equation}\label{BPField}
\begin{split}
& E_x = \mathcal{E}_0\mathrm{e}^{-\left(\sqrt{2\ln(2)}\frac{(t-z)}{\tau}\right)^2}\mathrm{e}^{-\frac{r^2}{w^2(z)}}
\left(S_1 + \epsilon^2\left[ \nu^2 S_3 -\frac{\rho^4S_4}{4} \right] \right. \\
& \qquad \qquad \qquad \qquad \left.+ \epsilon^4\left[ \frac{S_3}{8} -\frac{\rho^2S_4}{4} -\frac{\rho^2(\rho^2-16\nu^2)S_5}{16} - \frac{\rho^4(\rho^2+2\nu^2)S_6}{8} +\frac{\rho^8S_7}{32}\right]+\mathcal{O}(\epsilon^6)\right), \\
& E_y = \mathcal{E}_0\nu\eta\mathrm{e}^{-\left(\sqrt{2\ln(2)}\frac{(t-z)}{\tau}\right)^2}\mathrm{e}^{-\frac{r^2}{w^2(z)}}
\left( \epsilon^2S_3 + \epsilon^4\left[ \rho^2S_5 - \frac{\rho^4S_6}{4} \right]+\mathcal{O}(\epsilon^5)\right),\\
& E_z =\mathcal{E}_0 \nu\mathrm{e}^{-\left(\sqrt{2\ln(2)}\frac{(t-z)}{\tau}\right)^2}\mathrm{e}^{-\frac{r^2}{w^2(z)}}
\left(\epsilon S_2 + \epsilon^3\left[ -\frac{ S_3}{2} +\rho^2 S_4 -\frac{\rho^4S_5}{4}\right] +\mathcal{O}(\epsilon^5) \right),
\end{split}
\end{equation}
\begin{equation}\label{BPFieldB}
\begin{split}
& B_x=0,\qquad B_z =\mathcal{E}_0 \eta\mathrm{e}^{-\left(\sqrt{2\ln(2)}\frac{(t-z)}{\tau}\right)^2}\mathrm{e}^{-\frac{r^2}{w^2(z)}}
\left(\epsilon S_2 + \epsilon^3\left[ \frac{ S_3}{2} +\frac{\rho^2 S_4}{2} -\frac{\rho^4S_5}{4}\right] +\mathcal{O}(\epsilon^5)\right)\\
& B_y = \mathcal{E}_0\mathrm{e}^{-\left(\sqrt{2\ln(2)}\frac{(t-z)}{\tau}\right)^2}\mathrm{e}^{-\frac{r^2}{w^2(z)}}
\left(S_1 + \epsilon^2\left[ \frac{\rho^2 S_3}{2} -\frac{\rho^4S_4}{4} \right] \right. \\
& \qquad \qquad \qquad \qquad \left.+ \epsilon^4\left[ -\frac{S_3}{8} +\frac{\rho^2S_4}{4} +\frac{5\rho^4S_5}{16} - \frac{\rho^6S_6}{4} +\frac{\rho^8S_7}{32}\right]+\mathcal{O}(\epsilon^6)\right).
\end{split}
\end{equation}
\end{widetext}
In these formulae $\nu=x/w_0$, $\eta=y/w_0$, $\rho^2 = \nu^2+\eta^2$ and 
\[
S_n = \left(\frac{1}{\sqrt{1+\zeta(z)^2}}\right)^n\mathrm{sin}\left[\Phi+(n-1) \ \arctan(\zeta)\right],
\] 
\[
C_n = \left(\frac{1}{\sqrt{1+\zeta(z)^2}}\right)^n\mathrm{cos}\left[\Phi+(n-1)\ \arctan(\zeta)\right],
\]where an explicit expression for $\Phi$ can be found in Eq.~\eqref{Phi}.
Moreover, the pulse energy calculated with accuracy up to the fourth order in $\epsilon$ reads
\begin{equation}\label{WBP}
W_{\mathrm{BPA}} \approx \frac{\mathcal{E}_0^2}{2} \frac{\pi w_0^2}{2}\left( 1+ \frac{\epsilon^2}{4} + \frac{\epsilon^4}{8}\right) \frac{\tau}{2}\sqrt{\frac{\pi}{\ln(2)}}.
\end{equation}



\begin{thebibliography}{58}
\expandafter\ifx\csname
natexlab\endcsname\relax\def\natexlab#1{#1}\fi
\expandafter\ifx\csname bibnamefont\endcsname\relax
  \def\bibnamefont#1{#1}\fi
\expandafter\ifx\csname bibfnamefont\endcsname\relax
  \def\bibfnamefont#1{#1}\fi
\expandafter\ifx\csname citenamefont\endcsname\relax
  \def\citenamefont#1{#1}\fi
\expandafter\ifx\csname url\endcsname\relax
  \def\url#1{\texttt{#1}}\fi
\expandafter\ifx\csname urlprefix\endcsname\relax\def\urlprefix{URL
}\fi \providecommand{\bibinfo}[2]{#2}
\providecommand{\eprint}[2][]{\url{#2}}

\bibitem{BreitWheeler}
\bibinfo{author}{\bibfnamefont{G.}~\bibnamefont{Breit}} \bibnamefont{and}
\bibinfo{author}{\bibfnamefont{J.~A.}~\bibnamefont{Wheeler}},
\emph{\bibinfo{Title}{Collision of Two Light Quanta}},
\bibinfo{journal}{Phys. Rev.} \pmb{\bibinfo{volume}{46}},
\bibinfo{pages}{1087} (\bibinfo{year}{1934}).

\bibitem{Reiss1}
\bibinfo{author}{\bibfnamefont{H.~R.}~\bibnamefont{Reiss}}, 
\emph{\bibinfo{Title}{Absorption of light by light}},
\bibinfo{journal}{J. Math. Phys.},
\pmb{\bibinfo{volume}{3}},
\bibinfo{pages}{59} (\bibinfo{year}{1962}).

  \bibitem{NikishovRitus}
 \bibinfo{author}{\bibfnamefont{A.~I.}~\bibnamefont{Nikishov}}\bibnamefont{ and}
\bibinfo{author}{\bibfnamefont{V.~I.}~\bibnamefont{Ritus}},
\emph{\bibinfo{Title}{Quantum processes in the field of a plane electromagnetic wave and in constant field I. }},
\bibinfo{journal}{Zh. Eksp. Teor. Fiz.} \pmb{\bibinfo{volume}{46}},
 (\bibinfo{year}{1963}).
 
 
  \bibitem{NikishovRitus2}
 \bibinfo{author}{\bibfnamefont{A.~I.}~\bibnamefont{Nikishov}}\bibnamefont{ and}
\bibinfo{author}{\bibfnamefont{V.~I.}~\bibnamefont{Ritus}},
\emph{\bibinfo{Title}{Pair production by a photon and photon emission by an electron in the field of an intense electromagnetic wave and in a constant field}},
\bibinfo{journal}{Zh. Eksp. Teor. Fiz.} \pmb{\bibinfo{volume}{52}},
\bibinfo{pages}{1707}, (\bibinfo{year}{1967}).

\bibitem{baier}
\bibinfo{author}{\bibfnamefont{V.~N.~}\bibnamefont{Ba\u{\i}er}},
\bibinfo{author}{\bibfnamefont{A.~I.~}\bibnamefont{Mil'shte\u{\i}n}}
\bibnamefont{and}
\bibinfo{author}{\bibfnamefont{V.~M.~}\bibnamefont{Strakhovenko}},
\emph{\bibinfo{title}{Interaction between a photon and an intense electromagnetic wave}},
\bibinfo{journal}{Zh. Eksp. Teo. Fiz.} \pmb{\bibinfo{volume}{69}},
\bibinfo{pages}{1893} (\bibinfo{year}{1975}); [\bibinfo{journal}{Sov. Phys. JETP} \textbf{\bibinfo{volume}{42}},
\bibinfo{pages}{961} (\bibinfo{year}{1976})].


\bibitem{Reiss}
\bibinfo{author}{\bibfnamefont{H.~R.}~\bibnamefont{Reiss}}, 
\emph{\bibinfo{Title}{Production of Electron Pairs from a Zero-Mass State}},
\bibinfo{journal}{Phys. Rev. Lett.},
\pmb{\bibinfo{volume}{26}},
\bibinfo{pages}{1072} (\bibinfo{year}{1971}).

\bibitem{RitusReview}
\bibinfo{author}{\bibfnamefont{V.~I.}~\bibnamefont{Ritus}}, 
\emph{\bibinfo{Title}{Quantum effects of the interaction of elementary particles with an intense electromagnetic field}},
\bibinfo{journal}{J. Sov. Laser Res.} \pmb{\bibinfo{volume}{6}},
\bibinfo{pages}{497} (\bibinfo{year}{1985}).
 
 \bibitem{Burke}
\bibinfo{author}{\bibfnamefont{D.~L.}~\bibnamefont{Burke}}\emph{\bibinfo{Title}{ et al.}},
\emph{\bibinfo{Title}{Positron production in multiphoton light-by-light scattering}},
\bibinfo{journal}{Phys. Rev. Lett.} \pmb{\bibinfo{volume}{79}},
\bibinfo{pages}{1626} (\bibinfo{year}{1997}).

\bibitem{STAR}
\bibinfo{author}{\bibfnamefont{J.}~\bibnamefont{Adam}} \emph{\bibinfo{Title}{et al.}} \bibinfo{author}{\bibnamefont{(STAR collaboration)}},
\emph{\bibinfo{Title}{Measurement of $e^+e^-$ Momentum and Angular Distributions from Linearly Polarized Photon Collisions}},
\bibinfo{journal}{Phys. Rev. Lett.} \pmb{\bibinfo{volume}{127}},
\bibinfo{pages}{052302} (\bibinfo{year}{2021}).

\bibitem{diPiazzaReview}
\bibinfo{author}{\bibfnamefont{A.}~\bibnamefont{Di Piazza}},
\bibinfo{author}{\bibfnamefont{C.}~\bibnamefont{M\"uller}}, \bibinfo{author}{\bibfnamefont{K.~Z.}~\bibnamefont{Hatsagortsyan}}
\bibnamefont{and} \bibinfo{author}{\bibfnamefont{C.~H.}~\bibnamefont{Keitel}},
\emph{\bibinfo{Title}{Extremely high-intensity laser interactions with fundamental quantum systems}},
\bibinfo{journal}{Rev. Mod. Phys.} \pmb{\bibinfo{volume}{84}},
\bibinfo{pages}{1177} (\bibinfo{year}{2012}).

\bibitem{Meuren2019}
\bibinfo{author}{\bibfnamefont{S.}~\bibnamefont{Meuren}},
\emph{\bibinfo{Title}{Probing strong-field QED at FACET-II (SLAC E-320)}},
\emph{\bibinfo{Title}{$https://conf.slac.stanford.edu/facet\\-2-2019/sites/facet-2-2019.conf.slac.stanford.edu\\files//basic-page-docs/sfqed_2019.pdf$}}.

\bibitem{LUXE}
\bibinfo{author}{\bibfnamefont{H.}~\bibnamefont{Abramowicz}}\emph{\bibinfo{Title}{ et al.}},
\emph{\bibinfo{Title}{Letter of Intent for the LUXE Experiment}},
\emph{\bibinfo{Title}{arXiv:1909.00860}}.

\bibitem{LUXE2}
\bibinfo{author}{\bibfnamefont{H.}~\bibnamefont{Abramowicz}}\emph{\bibinfo{Title}{ et al.}},
\emph{\bibinfo{Title}{Conceptual design report for the LUXE experiment}},
\bibinfo{journal}{Eur. Phys. J. Spec. Top.} \pmb{\bibinfo{volume}{230}},
\bibinfo{pages}{2445}, (\bibinfo{year}{2021}). 

\bibitem{Appleton}
\bibinfo{author}{\bibfnamefont{C.~H.}~\bibnamefont{Keitel}} \emph{et al.},
\emph{\bibinfo{Title}{Photo-induced pair production and strong field QED on Gemini}},
\emph{\bibinfo{Title}{arXiv:2103.06.059}}.

\bibitem{Salgado}
\bibinfo{author}{\bibfnamefont{F.~C.}~\bibnamefont{Salgado}}\emph{\bibinfo{Title}{ et al.}},
\emph{\bibinfo{Title}{Towards pair production in the non-perturbative regime}},
\bibinfo{journal}{New J. Phys.} \pmb{\bibinfo{volume}{21}},
\bibinfo{pages}{105002}, (\bibinfo{year}{2021}). 

\bibitem{Heinzl}
\bibinfo{author}{\bibfnamefont{T.}~\bibnamefont{Heinzl}},
\bibinfo{author}{\bibfnamefont{A.}~\bibnamefont{Ilderton}} \bibnamefont{and}
\bibinfo{author}{\bibfnamefont{M.}~\bibnamefont{Marklund}},
\emph{\bibinfo{Title}{Finite size effects in stimulated laser pair production}},
\bibinfo{journal}{Phys. Lett. B} \pmb{\bibinfo{volume}{692}},
\bibinfo{pages}{250} (\bibinfo{year}{2010}).

\bibitem{Krajewska2012}
\bibinfo{author}{\bibfnamefont{K.}~\bibnamefont{Krajewska}} 
\bibnamefont{and} \bibinfo{author}{\bibfnamefont{J.~Z.}~\bibnamefont{Kami\'nski}},
\emph{\bibinfo{Title}{Breit-Wheeler process in intense short laser pulses}},
\bibinfo{journal}{Phys. Rev. A} \pmb{\bibinfo{volume}{86}},
\bibinfo{pages}{052104} (\bibinfo{year}{2012}).

\bibitem{Kaempfer2012}
\bibinfo{author}{\bibfnamefont{A.~I.}~\bibnamefont{Titov}},
\bibinfo{author}{\bibfnamefont{H.}~\bibnamefont{Takabe}},
\bibinfo{author}{\bibfnamefont{B.}~\bibnamefont{K\"ampfer}} \bibnamefont{and}
\bibinfo{author}{\bibfnamefont{A.}~\bibnamefont{Hosaka}},
\emph{\bibinfo{Title}{Enhanced subthreshold e+e- production in short laser pulses}},
\bibinfo{journal}{Phys. Rev. Lett.} \pmb{\bibinfo{volume}{108}},
\bibinfo{pages}{240406} (\bibinfo{year}{2012}).

\emph{\bibinfo{Title}{Breit-Wheeler process in very short electromagnetic pulses}},
\bibinfo{journal}{Phys. Rev. A} \pmb{\bibinfo{volume}{87}},
\bibinfo{pages}{042106} (\bibinfo{year}{2013}).

\bibitem{Jansen2013}
\bibinfo{author}{\bibfnamefont{M.~J.~A.}~\bibnamefont{Jansen}}, 
\bibnamefont{and} \bibinfo{author}{\bibfnamefont{C.}~\bibnamefont{M\"uller}},
\emph{\bibinfo{Title}{Strongly enhanced pair production in combined high- and low-frequency laser fields}},
\bibinfo{journal}{Phys. Rev. A} \pmb{\bibinfo{volume}{88}},
\bibinfo{pages}{052125} (\bibinfo{year}{2013}).

\bibitem{VillalbaChavez:2012bb}
\bibinfo{author}{\bibfnamefont{S.}~\bibnamefont{Villalba-Chavez}}
\bibnamefont{and}
\bibinfo{author}{\bibfnamefont{C.} \bibnamefont{M\"uller}}.
\emph{\bibinfo{title}{Photo-production of scalar particles in the field of a circularly polarized laser beam}},
\bibinfo{journal}{Phys. Lett. B},
\textbf{\bibinfo{volume}{718}},
\bibinfo{pages}{992} (\bibinfo{year}{2013}).

\bibitem{Krajewska2014}
\bibinfo{author}{\bibfnamefont{K.}~\bibnamefont{Krajewska}} 
\bibnamefont{and} \bibinfo{author}{\bibfnamefont{J.~Z.}~\bibnamefont{Kami\'nski}},
\emph{\bibinfo{Title}{Coherent combs of antimatter from nonlinear electron-positron-pair creation}},
\bibinfo{journal}{Phys. Rev. A} \pmb{\bibinfo{volume}{90}},
\bibinfo{pages}{052108} (\bibinfo{year}{2014}).

\bibitem{Meuren:2014uia}
\bibinfo{author}{\bibfnamefont{S.}~\bibnamefont{Meuren}}, 
\bibinfo{author}{\bibfnamefont{K.~Z.}~\bibnamefont{Hatsagortsyan}},
\bibinfo{author}{\bibfnamefont{C.~H.}~\bibnamefont{Keitel}}  \bibnamefont{and}
\bibinfo{author}{\bibfnamefont{A.~Di}~\bibnamefont{Piazza}}, 
\emph{\bibinfo{title}{Polarization-operator approach to pair creation in short laser pulses}},
\bibinfo{journal}{ Phys.\ Rev.\ D} \textbf{\bibinfo{volume}{91}},
\bibinfo{pages}{013009} (\bibinfo{year}{2015}).

\bibitem{Jansen2017}
\bibinfo{author}{\bibfnamefont{M.~J.~A.}~\bibnamefont{Jansen}} \bibnamefont{and}
\bibinfo{author}{\bibfnamefont{C.}~\bibnamefont{M\"uller}},
\emph{\bibinfo{Title}{Strong-field Breit–Wheeler pair production in two consecutive laser pulses with variable time delay}},
\bibinfo{journal}{Phys. Lett. B} \pmb{\bibinfo{volume}{766}},
\bibinfo{pages}{71} (\bibinfo{year}{2017}).

\bibitem{Kaempfer2018}
\bibinfo{author}{\bibfnamefont{A.~I.}~\bibnamefont{Titov}},
\bibinfo{author}{\bibfnamefont{H.}~\bibnamefont{Takabe}} \bibnamefont{and}
\bibinfo{author}{\bibfnamefont{B.}~\bibnamefont{K\"ampfer}},
\emph{\bibinfo{Title}{ Breit-Wheeler process in short laser double pulses}},
\bibinfo{journal}{Phys. Rev. D} \pmb{\bibinfo{volume}{98}},
\bibinfo{pages}{036022} (\bibinfo{year}{2018}).

\bibitem{Grobe2018}
\bibinfo{author}{\bibfnamefont{Q.~Z.}~\bibnamefont{Lv}},
\bibinfo{author}{\bibfnamefont{S.}~\bibnamefont{Dong}}, \bibinfo{author}{\bibfnamefont{Y. T.}~\bibnamefont{Li}},
\bibinfo{author}{\bibfnamefont{Z. M.}~\bibnamefont{Sheng}}, \bibinfo{author}{\bibfnamefont{Q.}~\bibnamefont{Su}} 
\bibnamefont{and} \bibinfo{author}{\bibfnamefont{R.}~\bibnamefont{Grobe}},
\emph{\bibinfo{Title}{Role of the spatial
inhomogeneity on the laser-induced vacuum decay}},
\bibinfo{journal}{Phys. Rev. A} \pmb{\bibinfo{volume}{97}},
\bibinfo{pages}{022515} (\bibinfo{year}{2018}).

\bibitem{Kaempfer2020}
\bibinfo{author}{\bibfnamefont{A.~I.}~\bibnamefont{Titov}} \bibnamefont{and}
\bibinfo{author}{\bibfnamefont{B.}~\bibnamefont{K\"ampfer}},
\emph{\bibinfo{Title}{ Nonlinear Breit-Wheeler process with linearly polarized beams}},
\bibinfo{journal}{Eur. Phys. J. D} \pmb{\bibinfo{volume}{74}},
\bibinfo{pages}{218} (\bibinfo{year}{2020}).

\bibitem{Tang}
\bibinfo{author}{\bibfnamefont{S.}~\bibnamefont{Tang}} 
\bibnamefont{and} \bibinfo{author}{\bibfnamefont{B.}~\bibnamefont{King}},
\emph{\bibinfo{Title}{Pulse envelope effects in nonlinear Breit-Wheeler pair creation}},
\bibinfo{journal}{Phys. Rev. D} \pmb{\bibinfo{volume}{104}},
\bibinfo{pages}{096019} (\bibinfo{year}{2021}).

\bibitem{X}
\bibinfo{author}{\bibfnamefont{Special aspects of Breit-Wheeler pair production are studied in:}}
\bibinfo{author}{\bibfnamefont{S.}~\bibnamefont{Meuren}}, \bibinfo{author}{\bibfnamefont{K.~Z.}~\bibnamefont{Hatsagortsyan}}, \bibinfo{author}{\bibfnamefont{C.~H.}~\bibnamefont{Keitel}} and \bibinfo{author}{\bibfnamefont{A. Di}~\bibnamefont{Piazza}}, 
\emph{\bibinfo{Title}{High-Energy Recollision Processes of Laser-Generated Electron-Positron Pairs}},
\bibinfo{journal}{Phys. Rev. Lett.} \pmb{\bibinfo{volume}{114}},
\bibinfo{pages}{143201} (\bibinfo{year}{2015});
\bibinfo{author}{\bibfnamefont{M.~J.~A.}~\bibnamefont{Jansen}},
\bibinfo{author}{\bibfnamefont{J.~Z.}~\bibnamefont{Kami\'nski}},
\bibinfo{author}{\bibfnamefont{K.}~\bibnamefont{Krajewska}} \bibnamefont{and}
\bibinfo{author}{\bibfnamefont{C.}~\bibnamefont{M\"uller}},
\emph{\bibinfo{Title}{Strong-field Breit-Wheeler pair production in short laser pulses: Relevance of spin effects}},
\bibinfo{journal}{Phys. Rev. D} \pmb{\bibinfo{volume}{94}},
\bibinfo{pages}{013010} (\bibinfo{year}{2016});
\bibinfo{author}{\bibfnamefont{T.}~\bibnamefont{Nousch}},
\bibinfo{author}{\bibfnamefont{D.}~\bibnamefont{Seipt}}, 
\bibinfo{author}{\bibfnamefont{B.}~\bibnamefont{K\"ampfer}} and \bibinfo{author}{\bibfnamefont{A.~I.}~\bibnamefont{Titov}}, 
\emph{\bibinfo{Title}{Spectral caustics in laser assisted Breit–Wheeler process}},
\bibinfo{journal}{Phys. Lett. B} \pmb{\bibinfo{volume}{755}},
\bibinfo{pages}{162} (\bibinfo{year}{2016});
\bibinfo{author}{\bibfnamefont{F.}~\bibnamefont{Wan}},
\bibinfo{author}{\bibfnamefont{Y.}~\bibnamefont{Wang}},
\bibinfo{author}{\bibfnamefont{R.~T.}~\bibnamefont{Guo}}, 
\bibinfo{author}{\bibfnamefont{R.~R.}~\bibnamefont{Chen}},
\bibinfo{author}{\bibfnamefont{R.}~\bibnamefont{Shaisultanov}},
\bibinfo{author}{\bibfnamefont{Z.~F.}~\bibnamefont{Xu}},
\bibinfo{author}{\bibfnamefont{K.~Z.}~\bibnamefont{Hatsagortsyan}}, 
\bibinfo{author}{\bibfnamefont{C.~H.}~\bibnamefont{Keitel}} and \bibinfo{author}{\bibfnamefont{J.~X.}~\bibnamefont{Li}}, 
\emph{\bibinfo{Title}{High-energy $\gamma$-photon polarization in
nonlinear Breit-Wheeler pair production and $\gamma$ polarimetry}},
\bibinfo{journal}{Phys. Rev. Research} \pmb{\bibinfo{volume}{2}},
\bibinfo{pages}{032049(R)} (\bibinfo{year}{2020});
\bibinfo{author}{\bibfnamefont{Y.}~\bibnamefont{Lu}},
\bibinfo{author}{\bibfnamefont{N.}~\bibnamefont{Christensen}}, 
\bibinfo{author}{\bibfnamefont{Q.}~\bibnamefont{Su}} \bibnamefont{and} \bibinfo{author}{\bibfnamefont{R.}~\bibnamefont{Grobe}},
\emph{\bibinfo{Title}{Space-time-resolved Breit-Wheeler process for a model system}},
\bibinfo{journal}{Phys. Rev. A} \pmb{\bibinfo{volume}{101}},
\bibinfo{pages}{022503} (\bibinfo{year}{2020});
\bibinfo{author}{\bibfnamefont{A.}~\bibnamefont{Golub}},
\bibinfo{author}{\bibfnamefont{S.}~\bibnamefont{Villalba-Ch\'avez}} \bibnamefont{and}
\bibinfo{author}{\bibfnamefont{C.}~\bibnamefont{M\"uller}},
\emph{\bibinfo{Title}{Strong-field Breit-Wheeler pair production in $\mathrm{QED}_{2+1}$}},
\bibinfo{journal}{Phys. Rev. D} \pmb{\bibinfo{volume}{103}},
\bibinfo{pages}{096002} (\bibinfo{year}{2021}).

\bibitem{diPiazza2016}
\bibinfo{author}{\bibfnamefont{A.}~\bibnamefont{Di Piazza}},
\emph{\bibinfo{Title}{Nonlinear Breit-Wheeler Pair Production in a Tightly Focused Laser Beam}},
\bibinfo{journal}{Phys. Rev. Lett.} \pmb{\bibinfo{volume}{117}},
\bibinfo{pages}{213201} (\bibinfo{year}{2016}).

\bibitem{diPiazza2021}
\bibinfo{author}{\bibfnamefont{A.}~\bibnamefont{Di Piazza}},
\emph{\bibinfo{Title}{WKB electron wave functions in a tightly focused laser beam}},
\bibinfo{journal}{Phys. Rev. D} \pmb{\bibinfo{volume}{103}},
\bibinfo{pages}{076011} (\bibinfo{year}{2021}).

\bibitem{Riconda}
\bibinfo{author}{\bibfnamefont{A.}~\bibnamefont{Mercuri-Baron}}\emph{\bibinfo{Title}{ et al.}},
\emph{\bibinfo{Title}{Impact of the laser spatio-temporal shape on
Breit–Wheeler pair production}},
\bibinfo{journal}{New J. Phys.} \pmb{\bibinfo{volume}{23}},
\bibinfo{pages}{085006}, (\bibinfo{year}{2021}).

\bibitem{Blackburn}
\bibinfo{author}{\bibfnamefont{T.~G.}~\bibnamefont{Blackburn}} \bibnamefont{and}
\bibinfo{author}{\bibfnamefont{M.}~\bibnamefont{Marklund}},
\emph{\bibinfo{Title}{Nonlinear Breit-Wheeler pair creation with bremsstrahlung $\gamma$ rays}},
\bibinfo{journal}{Plasma Phys. Control. Fusion} \pmb{\bibinfo{volume}{60}},
\bibinfo{pages}{054009}, (\bibinfo{year}{2018}).

\bibitem{Hartin}
\bibinfo{author}{\bibfnamefont{A.}~\bibnamefont{Hartin}},
\bibinfo{author}{\bibfnamefont{A.}~\bibnamefont{Ringwald}} \bibnamefont{and}
\bibinfo{author}{\bibfnamefont{N.}~\bibnamefont{Tapia}},
\emph{\bibinfo{Title}{Measuring the boiling point of the vacuum of quantum electrodynamics}},
\bibinfo{journal}{Phys. Rev. D} \pmb{\bibinfo{volume}{99}},
\bibinfo{pages}{036008} (\bibinfo{year}{2019}).

 \bibitem{Eckey}
\bibinfo{author}{\bibfnamefont{A.}~\bibnamefont{Eckey}},
\bibinfo{author}{\bibfnamefont{A.~B.}~\bibnamefont{Voitkiv}} \bibnamefont{and}
\bibinfo{author}{\bibfnamefont{C.}~\bibnamefont{M\"uller}},
\emph{\bibinfo{Title}{Strong-field Breit-Wheeler pair production with bremsstrahlung gamma-rays in the perturbative-to-nonperturbative transition regime}},  
\bibinfo{journal}{Phys. Rev. A},
\pmb{\bibinfo{volume}{105}},
\bibinfo{pages}{013105} (\bibinfo{year}{2022}).

 \bibitem{Golub}
\bibinfo{author}{\bibfnamefont{A.}~\bibnamefont{Golub}},
\bibinfo{author}{\bibfnamefont{S.}~\bibnamefont{Villalba-Ch\'avez}},
\bibinfo{author}{\bibfnamefont{H.}~\bibnamefont{Ruhl}} \bibnamefont{and}
\bibinfo{author}{\bibfnamefont{C.}~\bibnamefont{M\"uller}},
\emph{\bibinfo{Title}{Linear Breit-Wheeler pair production by high-energy bremsstrahlung photons colliding with an intense x-ray laser pulse}},
\bibinfo{journal}{Phys. Rev. D} \pmb{\bibinfo{volume}{103}},
\bibinfo{pages}{016009} (\bibinfo{year}{2021}).

\bibitem{King2015}
\bibinfo{author}{\bibfnamefont{C. N.}~\bibnamefont{Harvey}},
\bibinfo{author}{\bibfnamefont{A.}~\bibnamefont{Ilderton}} and \bibinfo{author}{\bibfnamefont{B.}~\bibnamefont{King}} 
\emph{\bibinfo{Title}{Testing numerical implementations of strong-field electrodynamics}},
\bibinfo{journal}{Phys. Rev. A} \pmb{\bibinfo{volume}{91}},
\bibinfo{pages}{013822} (\bibinfo{year}{2015}).

\bibitem{Meuren}
\bibinfo{author}{\bibfnamefont{A.}~\bibnamefont{Di Piazza}}, 
\bibinfo{author}{\bibfnamefont{M.}~\bibnamefont{Tamburini}},
 \bibinfo{author}{\bibfnamefont{S.}~\bibnamefont{Meuren}} \bibnamefont{and}
 \bibinfo{author}{\bibfnamefont{C.~H.}~\bibnamefont{Keitel}},
\emph{\bibinfo{Title}{Implementing nonlinear Compton scattering beyond the local-constant-field approximation}},
\bibinfo{journal}{Phys. Rev. A} \pmb{\bibinfo{volume}{98}},
\bibinfo{pages}{012134}, (\bibinfo{year}{2018}).

\bibitem{King2019}
\bibinfo{author}{\bibfnamefont{A.}~\bibnamefont{Ilderton}}, \bibinfo{author}{\bibfnamefont{B.}~\bibnamefont{King}} and
\bibinfo{author}{\bibfnamefont{D.}~\bibnamefont{Seipt}},
\emph{\bibinfo{Title}{Extended locally constant field approximation for nonlinear Compton scattering}},
\bibinfo{journal}{Phys. Rev. A} \pmb{\bibinfo{volume}{99}},
\bibinfo{pages}{042121} (\bibinfo{year}{2019}).

\bibitem{King2020}
\bibinfo{author}{\bibfnamefont{B.}~\bibnamefont{King}},
\emph{\bibinfo{Title}{Uniform locally constant field approximation for photon-seeded pair production}},
\bibinfo{journal}{Phys. Rev. A} \pmb{\bibinfo{volume}{101}},
\bibinfo{pages}{042508} (\bibinfo{year}{2020}).
 
\bibitem{King2021}
\bibinfo{author}{\bibfnamefont{D.}~\bibnamefont{Seipt}} 
\bibnamefont{and} \bibinfo{author}{\bibfnamefont{B.}~\bibnamefont{King}},
\emph{\bibinfo{Title}{Spin- and polarization-dependent locally-constant-field-approximation rates for nonlinear Compton and Breit-Wheeler processes}},
\bibinfo{journal}{Phys. Rev. A} \pmb{\bibinfo{volume}{102}},
\bibinfo{pages}{052805} (\bibinfo{year}{2020}).

\bibitem{Esarey1}
\bibinfo{author}{\bibfnamefont{E.}~\bibnamefont{Esarey}}, \bibinfo{author}{\bibfnamefont{P.}~\bibnamefont{Sprangle}}, \bibinfo{author}{\bibfnamefont{J.}~\bibnamefont{Krall}} and
\bibinfo{author}{\bibfnamefont{A.}~\bibnamefont{Ting}},
\emph{\bibinfo{Title}{Overview of plasma-based accelerator concepts}},
\bibinfo{journal}{IEEE  Trans.  Plasma.  Sci.} \pmb{\bibinfo{volume}{24}},
\bibinfo{pages}{252} (\bibinfo{year}{1996}).

\bibitem{Esarey2}
\bibinfo{author}{\bibfnamefont{E.}~\bibnamefont{Esarey}}, \bibinfo{author}{\bibfnamefont{C.~B.}~\bibnamefont{Schroeder}} and
\bibinfo{author}{\bibfnamefont{W.~P.}~\bibnamefont{Leemans}},
\emph{\bibinfo{Title}{Physics of laser-driven plasma-based electron accelerators}},
\bibinfo{journal}{Rev. Mod. Phys.} \pmb{\bibinfo{volume}{81}},
\bibinfo{pages}{1229} (\bibinfo{year}{2009}).

\bibitem{Lobet}
\bibinfo{author}{\bibfnamefont{M.}~\bibnamefont{Lobet}},
\bibinfo{author}{\bibfnamefont{X.}~\bibnamefont{Davoine}},
\bibinfo{author}{\bibfnamefont{E.}~\bibnamefont{d'Humières}} 
\bibnamefont{and} \bibinfo{author}{\bibfnamefont{L.}~\bibnamefont{Gremillet}},
\emph{\bibinfo{Title}{Generation of high-energy electron-positron pairs in the collision
of a laser-accelerated electron beam with a multipetawatt laser}},
\bibinfo{journal}{Phys. Rev. Accel. Beams} \pmb{\bibinfo{volume}{20}},
\bibinfo{pages}{043401} (\bibinfo{year}{2017}).

\bibitem{PartPhysGroup}
\bibinfo{author}{\bibfnamefont{P.~A.}~\bibnamefont{Zyla}}\emph{\bibinfo{Title}{ et al.}}\bibinfo{author}{\bibfnamefont{(Particle Data Group)}},
\emph{\bibinfo{Title}{Review of Particle Physics}},
\bibinfo{journal}{Prog. Theor. Exp. Phys.} \pmb{\bibinfo{volume}{8}} (\bibinfo{year}{2020}). 

\bibitem{Tsai}
\bibinfo{author}{\bibfnamefont{Y.-S.}~\bibnamefont{Tsai}},
\emph{\bibinfo{Title}{Pair production and bremsstrahlung of charged leptons}},
\bibinfo{journal}{Rev. Mod. Phys.} \pmb{\bibinfo{volume}{46}},
 \bibinfo{pages}{815} (\bibinfo{year}{1974}).
 
 \bibitem{Salamin}
\bibinfo{author}{\bibfnamefont{Y.~I.}~\bibnamefont{Salamin}},
\emph{\bibinfo{Title}{Fields of Gaussian beam beyond paraxial approximation}},
\bibinfo{journal}{Appl. Phys. B} \pmb{\bibinfo{volume}{86}} (\bibinfo{year}{2007}). 

 \bibitem{NIST}
\bibinfo{author}{\bibfnamefont{F.~W.~J.}~\bibnamefont{Olver}},
\bibinfo{author}{\bibfnamefont{D.~W.}~\bibnamefont{Lozier}},
\bibinfo{author}{\bibfnamefont{R.~F.}~\bibnamefont{Boisvert}}  \bibnamefont{and}
\bibinfo{author}{\bibfnamefont{C.~W.}~\bibnamefont{Clark}}, 
\emph{\bibinfo{Title}{NIST Handbook of Mathematical Functions}},
\bibinfo{publisher}{Cambridge University Press} (\bibinfo{year}{2010}). 

\bibitem{SuperGaussian}
\bibinfo{author}{\bibfnamefont{K.}~\bibnamefont{Gillen{-}Christandl}},
\bibinfo{author}{\bibfnamefont{G.~D.}~\bibnamefont{Gillen}},
\bibinfo{author}{\bibfnamefont{M.~J.}~\bibnamefont{Piotrowicz}} \bibnamefont{and}
\bibinfo{author}{\bibfnamefont{M.}~\bibnamefont{Saffman}},
\emph{\bibinfo{Title}{Comparison of Gaussian and super Gaussian laser beams
for addressing atomic qubits}},
\bibinfo{journal}{Appl. Phys. B} \pmb{\bibinfo{volume}{122}},
\bibinfo{pages}{131} (\bibinfo{year}{2016}). 
 
\bibitem{SGFG}
\bibinfo{author}{\bibfnamefont{M.}~\bibnamefont{Santarsiero}} \bibnamefont{and}
\bibinfo{author}{\bibfnamefont{R.}~\bibnamefont{Borghi}},
\emph{\bibinfo{Title}{Correspondence between super-Gaussian
and flattened Gaussian beams}},
\bibinfo{journal}{J. Opt. Soc. Am. A} \pmb{\bibinfo{volume}{16}},
\bibinfo{pages}{188} (\bibinfo{year}{1999}). 

 \bibitem{Bulanov}
\bibinfo{author}{\bibfnamefont{S.~S.}~\bibnamefont{Bulanov}},
\bibinfo{author}{\bibfnamefont{V.~D.}~\bibnamefont{Mur}},
\bibinfo{author}{\bibfnamefont{N.~B.}~\bibnamefont{Narozhny}},
\bibinfo{author}{\bibfnamefont{J.}~\bibnamefont{Nees}} \bibnamefont{and}
\bibinfo{author}{\bibfnamefont{V.~S.}~\bibnamefont{Popov}},
\emph{\bibinfo{Title}{Multiple Colliding Electromagnetic Pulses: A Way to Lower the Threshold of $e^+e^-$ Pair Production from Vacuum}},
\bibinfo{journal}{Phys. Rev. Lett.} \pmb{\bibinfo{volume}{104}},
\bibinfo{pages}{220404} (\bibinfo{year}{2010}). 

 \bibitem{ReportFedotov}
\bibinfo{author}{\bibfnamefont{A.}~\bibnamefont{Fedotov}},
\bibinfo{author}{\bibfnamefont{A.}~\bibnamefont{Ilderton}},
\bibinfo{author}{\bibfnamefont{F.}~\bibnamefont{Karbstein}},
\bibinfo{author}{\bibfnamefont{B.}~\bibnamefont{King}},
\bibinfo{author}{\bibfnamefont{D.}~\bibnamefont{Seipt}}, 
\bibinfo{author}{\bibfnamefont{H.}~\bibnamefont{Taya}} \bibnamefont{and}
\bibinfo{author}{\bibfnamefont{G.}~\bibnamefont{Torgrimsson}},
\emph{\bibinfo{Title}{Advances in QED with intense background fields}},
\emph{\bibinfo{Title}{$https://arxiv.org/pdf/2203.00019.pdf$}}.

\bibitem{NikishovRitus3}
\bibinfo{author}{\bibfnamefont{V.~I.}~\bibnamefont{Ritus}}, 
\emph{\bibinfo{Title}{Radiative effects and their enhancement in an intense electromagnetic field}},
\bibinfo{journal}{Zh. Eksp. Teor. Fis.} \pmb{\bibinfo{volume}{57}},
\bibinfo{pages}{2176} (\bibinfo{year}{1969});
[\bibinfo{journal}{J. Exp. Theor. Phys.} \pmb{\bibinfo{volume}{30}},
\bibinfo{pages}{1182} (\bibinfo{year}{1970}).]

\bibitem{Narozhny}
\bibinfo{author}{\bibfnamefont{N.~B.}~\bibnamefont{Narozhny}}, 
\emph{\bibinfo{Title}{Expansion parameter of perturbation theory in intense-field quantum electrodynamics}},
\bibinfo{journal}{Phys. Rev. D} \pmb{\bibinfo{volume}{21}},
\bibinfo{pages}{1176} (\bibinfo{year}{1980}).

\bibitem{Baumann}
\bibinfo{author}{\bibfnamefont{C.}~\bibnamefont{Baumann}}, 
\bibinfo{author}{\bibfnamefont{E. N.}~\bibnamefont{Nerush}},
\bibinfo{author}{\bibfnamefont{A.}~\bibnamefont{Pukhov}} \bibnamefont{and}
\bibinfo{author}{\bibfnamefont{I. Yu.}~\bibnamefont{Kostyukov}}, 
\emph{\bibinfo{Title}{Probing non-perturbative QED with electron-laser collisions}},
\bibinfo{journal}{Sci. Rep.} \pmb{\bibinfo{volume}{9}},
\bibinfo{pages}{9407} (\bibinfo{year}{2019}).


\bibitem{Podszus}
\bibinfo{author}{\bibfnamefont{T.}~\bibnamefont{Podszus}} \bibnamefont{and}
\bibinfo{author}{\bibfnamefont{A.}~\bibnamefont{Di Piazza}},
\emph{\bibinfo{Title}{High-energy behavior of strong-field QED in an intense plane wave}},
\bibinfo{journal}{Phys. Rev. D} \pmb{\bibinfo{volume}{99}},
\bibinfo{pages}{076004} (\bibinfo{year}{2019}).

\bibitem{Ilderton}
\bibinfo{author}{\bibfnamefont{A.}~\bibnamefont{Ilderton}},  
\emph{\bibinfo{Title}{Note on the conjectured breakdown of QED perturbation theory in strong fields}},
\bibinfo{journal}{Phys. Rev. D} \pmb{\bibinfo{volume}{99}},
\bibinfo{pages}{085002} (\bibinfo{year}{2019}).


\bibitem{Mironov}
\bibinfo{author}{\bibfnamefont{A.~A.}~\bibnamefont{Mironov}},  
\bibinfo{author}{\bibfnamefont{S.}~\bibnamefont{Meuren}} \bibnamefont{and}
\bibinfo{author}{\bibfnamefont{A.~M.}~\bibnamefont{Fedotov}},
\emph{\bibinfo{Title}{Resummation of QED radiative corrections in a strong constant crossed field}},
\bibinfo{journal}{Phys. Rev. D} \pmb{\bibinfo{volume}{102}},
\bibinfo{pages}{053005} (\bibinfo{year}{2020}).


\bibitem{Neville}
\bibinfo{author}{\bibfnamefont{R.~A.}~\bibnamefont{Neville}}  \bibnamefont{and}
\bibinfo{author}{\bibfnamefont{F.}~\bibnamefont{Rohrlich}},  
\emph{\bibinfo{Title}{Quantum Electrodynamics on Null Planes and Applications to Lasers}},
\bibinfo{journal}{Phys. Rev. D} \pmb{\bibinfo{volume}{3}},
\bibinfo{pages}{1692} (\bibinfo{year}{1971}). 

\bibitem{Mitter}
\bibinfo{author}{\bibfnamefont{W.}~\bibnamefont{Becker}}  \bibnamefont{and}
\bibinfo{author}{\bibfnamefont{H.}~\bibnamefont{Mitter}},  
\emph{\bibinfo{Title}{Vacuum polarization in laser fields}},
\bibinfo{journal}{J. Phys. A} \pmb{\bibinfo{volume}{8}},
\bibinfo{pages}{1638} (\bibinfo{year}{1975}). 

\bibitem{Leemanns2019}
\bibinfo{author}{\bibfnamefont{A.~J.}~\bibnamefont{Gonsalves}}\emph{\bibinfo{Title}{ et al.}},
\emph{\bibinfo{Title}{Petawatt Laser Guiding and Electron Beam Acceleration to 8 GeV
in a Laser-Heated Capillary Discharge Waveguide}},
\bibinfo{journal}{Phys. Rev. Lett.} \pmb{\bibinfo{volume}{122}},
\bibinfo{pages}{084801} (\bibinfo{year}{2019}).

\bibitem{Karsch}
\bibinfo{author}{\bibfnamefont{G.}~\bibnamefont{G\"otzfried}}\emph{\bibinfo{Title}{ et al.}},
\emph{\bibinfo{Title}{Physics of nanocoulomb-class electron beams in laser-plasma wakefields}},
\bibinfo{journal}{Phys. Rev. X} \pmb{\bibinfo{volume}{10}},
\bibinfo{pages}{041015} (\bibinfo{year}{2020}).


 


\end{thebibliography}
\end{document}